\documentclass[aps,prd,amssymb,amsmath,twocolumn,superscriptaddress]{revtex4-1}

\usepackage{graphicx,color}  
\usepackage{dcolumn}   
\usepackage{bm}        

\usepackage[english]{babel}
\usepackage[titletoc,title]{appendix}

\usepackage{mathtools}

\def\etal{{\it et al.\ }}

\def\ala{{\it \`a la\ }}

\newcommand{\wa}{{{\color{white}a}}}
\newcommand{\wg}{{{\color{white}g}}}

\allowdisplaybreaks

\newcommand{\rd}[1]{\frac{\rmd #1}{\rmd r}}

\newcommand{\tA}{\text{A}}
\newcommand{\tB}{\text{B}}
\newcommand{\tC}{\text{C}}
\newcommand{\tD}{\text{D}}
\newcommand{\tE}{\text{E}}
\newcommand{\tF}{\text{F}}
\newcommand{\tG}{\text{G}}
\newcommand{\tH}{\text{H}}
\newcommand{\tJ}{\text{J}}
\newcommand{\tK}{\text{K}}

\newcommand{\st}[1]{\widehat{\text{#1}}_{\text{L},\tS}}

\newcommand{\tP}{\text{P}}
\newcommand{\tQ}{\text{Q}}
\newcommand{\tR}{\text{R}}
\newcommand{\tS}{\text{S}}

\newcommand{\tW}{\text{W}}
\newcommand{\tZ}{\text{Z}}
\newcommand{\liexis}{\pounds_{\xi^\text{RW,S}}g_{ab}}
\newcommand{\liexir}{\pounds_{\xi^\text{RW,C}}g_{ab}}

\newcommand{\fF}{\mathcal{F}}
\newcommand{\tL}{\text{L}}
\newcommand{\tLL}{\text{LL}}

\newcommand{\NAF}{\text{NAF}}
\newcommand{\hb}[1]{\bar{h}^{(#1)}}

\newcommand{\beq}{\begin{equation}}
\newcommand{\eeq}{\end{equation}}

\newcommand{\rmd}{\mathrm{d}}

\newcommand{\tabdat}[2]{\(#1\times 10^{#2}\)}

\makeatletter
\newcommand{\subalign}[1]{%
  \vcenter{%
    \Let@ \restore@math@cr \default@tag
    \baselineskip\fontdimen10 \scriptfont\tw@
    \advance\baselineskip\fontdimen12 \scriptfont\tw@
    \lineskip\thr@@\fontdimen8 \scriptfont\thr@@
    \lineskiplimit\lineskip
    \ialign{\hfil$\m@th\scriptstyle##$&$\m@th\scriptstyle{}##$\crcr
      #1\crcr
    }%
  }
}
\makeatother

\begin{document}

\title{Gravitational Self-Force Regularization in the Regge-Wheeler and Easy Gauges}

\author{Jonathan E. Thompson}
\affiliation{Department of Physics, P.O. Box 118440, University of Florida, Gainesville, Florida 32611-8440}
\author{Barry Wardell}
\affiliation{School of Mathematical Sciences and Complex \& Adaptive Systems Laboratory,\\
University College Dublin, Belfield, Dublin 4, Ireland}
\author{Bernard F. Whiting}
\affiliation{Department of Physics, P.O. Box 118440, University of Florida, Gainesville, Florida 32611-8440}

\date{\today}

\begin{abstract}
We present numerical results for the gravitational self-force and redshift invariant calculated in the Regge-Wheeler and Easy gauges for circular orbits in a Schwarzschild background, utilizing the regularization framework introduced by Pound, Merlin, and Barack. The numerical calculation is performed in the frequency domain and requires the integration of a single second-order ODE, greatly improving computation times over more traditional Lorenz gauge numerical methods. A sufficiently high-order, analytic expansion of the Detweiler-Whiting singular field is gauge-transformed to both the Regge-Wheeler and Easy gauges and used to construct tensor-harmonic mode-sum regularization parameters.  We compare our results to the gravitational self-force calculated in the Lorenz gauge by explicitly gauge-transforming the Lorenz gauge self-force to the Regge-Wheeler and Easy gauges, and find that our results agree to a relative accuracy of~\(10^{-15}\) for an orbital radius of \(r_0=6M\) and~\(10^{-16}\) for an orbital radius of \(r_0=10M\).
\end{abstract}

\pacs{}

\maketitle

\section{Introduction}

Recent successes of the LIGO Scientific Collaboration to directly detect gravitational radiation \cite{LIGO1,LIGO2,LIGO3,LIGO4,LIGO5,LIGO6} have boosted interests in gravitational wave astrophysics. With the proposed launch date for the satellite-based LISA mission \cite{LISA} steadily approaching, source modeling efforts are rapidly progressing to build waveform models for candidate LISA sources.  One important candidate signal for the LISA mission is expected to arrive from the extreme mass-ratio inspiral (EMRI) of approximately solar-mass compact objects into supermassive black holes. Such systems will produce signals that remain in the detector for lengthy time periods, requiring highly precise models to extract accurate physical parameters from the data \cite{kipp}. One important effect to consider is the interaction of the compact object in the EMRI with its own gravitational field, the gravitational self-force, as these lengthy time periods generally extend into the radiation-reaction timescale \cite{LISAsymp}.

The formulation of the gravitational self-force within black hole perturbation theory has its foundational roots stemming from the works of Mino, Sasaki, and Tanaka \cite{MinoPRD55} and Quinn and Wald \cite{QuinnPRD56}, who separately introduced an expression for the self-force to first-order in the mass-ratio of the compact object (modeled by a point particle) to the supermassive black hole; the outcome of this formulation of the self-force is referred to as the MiSaTaQuWa equation. Alternative (and in some cases equivalent) regularization schemes such as mode-sum and zeta function regularization were subsequently proposed to remove the singularities introduced to the force by the point-particle source \cite{BarackPRD61,LoustoPRL84}. Further work by Detweiler and Whiting~\cite{DetWhitPRD67} allowed for a regularization scheme designed around the separation of the metric perturbation into singular and regular pieces, with the singular contributions physically motivated and akin to the Coulomb field of a point charge in electrodynamics.

Historically, the choice of Lorenz gauge in perturbation theory has been tightly linked with self-force calculations. This gauge choice is well motivated; the Lorenz-gauge field equations are manifestly hyperbolic, and the local expression of the particle's self-field assumes an isotropic form \cite{BarackPRD64}. Unfortunately, numerical integration of the Einstein field equations in the Lorenz gauge is non-trivial, as the field equations do not decouple and the numerics are complicated by gauge instabilities \cite{BarackPRD66,BarackPRD75}. More recent work has extended self-force regularization procedures to the radiation gauge by adjusting the standard Lorenz-gauge regularization scheme to accomodate string singularities present in the radiation gauge metric perturbation \cite{KeidlPRD82,ShahPRD83,ShahPRD86,PoundPRD89}.

One might ask whether it is possible to calculate the gravitational self-force in gauges common to the study of Schwarzschild black hole perturbations, such as the Regge-Wheeler (RW) gauge \cite{ReggePHR108} or the similar Easy (EZ) gauge recently introduced in \cite{TCW}. These gauge choices allow for fast and efficient reconstruction of the retarded metric perturbation generated by a point particle. Early work on this problem recovered the self-force for radially-infalling trajectories in the RW gauge \cite{LoustoPRL84}, but follow-up analysis showed that the singular contributions to the self-force in the RW gauge are not adequately regularized by standard Lorenz-gauge regularization techniques \cite{BarackPRD64}. Regularization of the RW gauge self-force using a tensor-harmonic decomposition of the local singular field was performed by Nakano \etal\cite{NakanoPRD68}, but only to first-order in a post-Newtonian expansion. These initial works in RW self-force regularization, along with advances in the understanding of how gauge choice affects regularization by Pound \etal\cite{PoundPRD89} and high-order tensor-harmonic expansion of the Detwieler-Whiting singular field by Wardell and Warburton \cite{WardellPRD92}, form the foundation for the work presented in this paper.

This paper is structured as follows. We review gravitational self-force regularization in Sec.~\ref{regscheme}, and demonstrate in Sec.~\ref{regularization} how the regularization is modified for the tensor-harmonic modes of the metric perturbation in the RW gauge. In Sec.~\ref{retsol} we review the gauge-invariant framework used to construct the retarded metric perturbation in the RW and EZ gauges. Special care is given in Sec.~\ref{l01retsol} to the low-multipole (\(\ell<2\)) modes of the retarded metric perturbation, which are calculated in the Zerilli gauge~\cite{ZerilliPRD2}. In Sec.~\ref{singfieldconst} we review the method used to construct the tensor-harmonic modes of the Detweiler-Whiting singular field introduced by Wardell and Warburton \cite{WardellPRD92}, and we outline the singular gauge transformation used to construct the singular field in the RW and EZ gauges. Finally, we present the numerical results in Sec.~\ref{results} for the regularized Detweiler redshift invariant and the gravitational self-force in both the RW and EZ gauges, and compare our results to the Lorenz gauge self-force through an explicit gauge transformation of the Lorenz gauge self-force.

We choose to work in geometrized units \(c=G=1\). The background Schwarzschild metric with mass \(M\) is labeled by \(g_{ab}\) in Schwarzschild coordinates \((t,r,\theta,\phi)\) with signature \((-,+,+,+)\). Lower-case Latin letters \(\{a,b,c,\dots\}\) indicate spacetime indices and Latin letters \(\{i,j,k,\dots\}\) indicate purely spatial indices, and we introduce \(f=1-2M/r\). We use the curvature conventions of Misner, Thorne, and Wheeler~\cite{MTW}. The symbol \(x\) denotes a spacetime event, and the subscript ``0'' indicates that a quantity is evaluated at the location of the point-particle perturbation, such that \(x_0=(0,r_0,\pi/2,0)\) and \(r_0\) is the constant orbital radius of the circular orbit.  The domain of integration is separated into two distinct regions, with the ``inner'' region \(r<r_0\) denoted by a ``\(-\)'' sign, and the ``outer'' region \(r>r_0\) denoted by a ``\(+\)'' sign.

We use an ``L'' to specify quantities calculated in the Lorenz gauge and an ``RW'' for the Regge-Wheeler and easy gauges, unless the distinction is important, in which case we explicitly write ``EZ'' for the easy gauge. Finally, for a continuous function \(F(r)\) with discontinuous derivative at \(r=r_0\), we write,
\begin{equation}
\left(\frac{\mathrm{d}F}{\mathrm{d}r}\right)_\pm\equiv\lim_{r\rightarrow r_0^{\pm}}\frac{\mathrm{d}F}{\mathrm{d}r}(r).
\end{equation}


\section{Self-Force Review \label{regscheme}}

We begin with a review of the perturbative analyses used to solve the Einstein field equations (EFEs) for a compact mass \(\mu\) in a circular orbit about a Schwarzschild black hole of mass \(M\), assuming \(\mu/M\ll1\). The physical spacetime metric is approximated as a background Schwarzschild metric plus a tensor perturbation, \(g_{ab}^\text{phys}=g_{ab}+h_{ab}\), and is a solution to the EFEs. When expanded to first-order in the mass-ratio \(\mu/M\), the EFEs take the form~\cite{BarackPRD75,DetweilerPRD77},
\beq
\label{linearizedEFE}
E_{ab}[h]=-16\pi T_{ab}+O(\mu^2/M^2),
\eeq
where we have introduced the linearized Einstein operator,
\pagebreak
\begin{align}
\nonumber E_{ab}[h]&=\nabla^c\nabla_ch_{ab}+\nabla_a\nabla_bh-2\nabla_{(a}\nabla^ch_{b)c}\\
\label{linearizedEO}&\quad+2R_a{}^c{}_b{}^dh_{cd}+g_{ab}\left(\nabla^c\nabla^dh_{cd}-\nabla^c\nabla_dh\right),
\end{align}
with \(h=g^{cd}h_{cd}\), and \(\nabla\) is the covariant derivative compatible with the background Schwarzschild metric.

The perturbing stress-energy of the compact mass is modeled as a point particle of mass \(\mu\) moving along a circular, equatorial geodesic of Schwarzschild spacetime, \(z^\mu(\tau)=\{t(\tau),r_0,\pi/2,\Omega\, t(\tau)\}\), where \(\tau\) denotes the particle's proper time and \(\Omega=\sqrt{M/r_0^3}\) is the frequency of the orbit. The stress-energy for the point particle is written,
\begin{align}
\nonumber T_{ab}&=\mu\int_{-\infty}^{\infty} \frac{u_au_b}{\sqrt{-g}}\,\delta^{4}[x^\mu-z^\mu(\tau)]\,\mathrm{d}\tau\\
\label{stressenergy}&=\mu\frac{u_au_b}{u^tr_0^2}\delta(r-r_0)\delta(\theta-\pi/2)\delta(\varphi-\Omega t),
\end{align}
with four-velocity \(u_a=(-\mathcal{E},0,0,\mathcal{L})\), and specific energy and angular momentum \(\mathcal{E}\) and \(\mathcal{L}\), respectively,
\begin{equation}
\label{energymomentum}
\mathcal{E}=\frac{r_0-2M}{\sqrt{r_0(r_0-3M)}}, \qquad \mathcal{L}=r_0\sqrt{\frac{M}{r_0-3M}}.
\end{equation}

The general force exerted by a vacuum perturbation \(h_{ab}\) on the compact mass is given by \cite{PoundPRD89},
\beq
\label{fullforce}
\mathcal{F}^a[h]=-\frac12\mu(g^{ab}+\tilde{u}^a\tilde{u}^b)(2\nabla_dh_{bc}-\nabla_bh_{cd})\tilde{u}^c\tilde{u}^d,
\eeq
written here as a vector field, where \(\tilde{u}^a\) is a smooth extension of the four-velocity off of the particle's worldline. To compute the self-force, each term of Eq.~\eqref{fullforce} is evaluated at the location of the particle. However, the RHS is formally singular at the location of the particle if one naively uses the metric perturbation arising from Eq.~\eqref{linearizedEFE}. This singularity in the force is not a physical result. Detweiler and Whiting \cite{DetWhitPRD67} find that the metric perturbation may be separated into \textit{singular} and \textit{regular} contributions,
\beq
h_{ab}=h^\tS_{ab}+h^\tR_{ab},
\eeq
such that each piece of the decomposition is individually a solution to Eq.~\eqref{linearizedEFE},
\begin{align}
E_{ab}[h^\tS]&=-16\pi T_{ab},\\
E_{ab}[h^\tR]&=0,
\end{align}
and \(h_{ab}^\tS\) does not contribute to the gravitational self-force, i.e.,
\beq
\label{forcehR}
\mathcal{F}^a_\text{self}=-\frac12\mu(g^{ab}+u^au^b)(2\nabla_dh^\tR_{bc}-\nabla_bh^\tR_{cd})u^cu^d.
\eeq
The quantities \(h_{ab}^\tS\) and \(h_{ab}^\tR\) are referred to as the Detweilier-Whiting singular and regular fields, respectively.

The Lorenz gauge is commonly used in gravitational self-force calculations. By introducing  the \textit{trace-reversed} metric perturbation \(\bar{h}_{ab}=h_{ab}-\frac12g_{ab}g^{cd}h_{cd}\), the Lorenz gauge condition is compactly written as,
\beq
\nabla^a\bar{h}_{ab}^\tL=0.
\eeq
In this gauge, the linearized Einstein operator in the EFEs reduces to a set of coupled wave equations acting on the trace-reversed metric components,
\beq
\label{LorenzEFE}
\nabla^c\nabla_c\bar{h}_{ab}^\tL+2R_a{}^c{}_b{}^d\bar{h}^\tL_{cd}=-16\pi T_{ab}.
\eeq
We assume that the metric perturbation for the remainder of this section is computed in the Lorenz gauge, and drop the ``L'' descriptor.

The retarded solution to Eq.~\eqref{LorenzEFE} can be found numerically, decomposed into a basis of scalar spherical harmonics,
\beq
\label{hscalar}
\bar{h}_{ab}^{\text{ret,}\hat{\ell}m}=Y_{\hat{\ell}m}(\theta,\phi)\int_0^{2\pi}\int_0^\pi \bar{h}_{ab}^\text{ret}Y^*_{\hat{\ell}m}(\theta',\phi')\,\rmd\Omega',
\eeq
with differential solid angle \(\rmd\Omega=\sin\theta\rmd\theta\rmd\phi\) and \(*\) the complex conjugation. Each \(\hat{\ell} m\)-mode of the Lorenz-gauge retarded metric perturbation \(h_{ab}^{\text{ret},\hat{\ell} m}\) is a finite \(C^0\) function of \(r\) at the location of the particle, but the infinite sum of the modes diverges as \(O(\hat{\ell})\). Furthermore, the \(\hat{\ell}m\)-modes of the force, which involve radial derivatives of the metric perturbation, have bounded jump-discontinuities at the particle. To calculate the regularized self-force,  the method of \textit{mode-sum regularization} was introduced by Barack and Ori \cite{BarackPRD61},
\beq
\label{modesumreg}
\fF^a_\text{self}=\sum_{\hat{\ell}=0}^\infty\left[\fF^{a,\hat{\ell}\pm}_\text{ret}-A^{a,\pm}\hat{L}-B^a-C^a/\hat{L}\right]-D^a,
\eeq
with \(\hat{L}=2\hat{\ell}+1\). The term \(\fF^{a,\hat{\ell}\pm}_\text{ret}\) is constructed from the scalar-harmonic modes of the retarded metric perturbation and evaluated at \(x_0\) in the inner or outer regions via the direction-dependent limit,
\begin{align}
\nonumber\fF_\text{ret}^{a,\hat{\ell}\pm}&=\lim_{r\rightarrow r_0^\pm}\sum_{m=-\hat{\ell}}^{\hat{\ell}}Y_{\hat{\ell}m}(\pi/2,0)\\
\label{eq:Fret}&\qquad\times\int_0^{2\pi}\int_0^\pi \fF^a[h^\text{ret}]Y^*_{\hat{\ell}m}(\theta',\phi')\,\rmd\Omega',
\end{align}
The quantities \(A^{a,\pm}\), \(B^a\), \(C^a\), and \(D^a\) are \textit{regularization parameters}, constants in \(\hat{\ell}\) derived from a local expansion of the singular field and known analytically in the Lorenz gauge for generic bound orbits of Schwarzschild \cite{BarackPRD61} and Kerr \cite{BarackPRD90} spacetimes. When subtracted mode-by-mode in Eq.~\eqref{modesumreg}, the \(\hat{\ell}\)-modes of the force fall off as \(O(\hat{\ell}^{-2})\), and the partial sums converge as \(O(\hat{\ell}^{-1})\). For circular orbits in Schwarzschild spacetime, the parameters \(A^{a,\pm}\) and \(B^a\) vanish for for all but the radial component of the force, and \(C^a=D^a=0\).

Instead of working in the scalar-harmonic \(\hat{\ell} m\) basis of Eq.~\eqref{hscalar}, one might choose to work in a \textit{tensor}-harmonic \(\ell m\) basis, such as the basis introduced for Lorenz-gauge self-force calculations by Barack and Lousto \cite{BarackPRD66} and Barack and Sago \cite{BarackPRD75} (that we shall refer to as the BLS basis). When decomposed into the BLS basis, the field equations separate into coupled scalar wave equations, allowing one to employ numerical methods developed for calculating the scalar self-force \cite{BarackPRD62,DetweilerPRD67}. To recover the \(\hat{\ell}\)-modes in Eq.~\eqref{eq:Fret}, one must re-project the tensor-harmonic \(\ell\)-modes onto the scalar-harmonic \(\hat{\ell}\)-modes, a process which generically requires the calculation of \(\ell+3\) tensor-harmonic modes \cite{BarackPRD75}. While relatively trivial for circular orbits in Schwarzschild spacetime, this re-projection becomes increasingly complicated and time-consuming when working on arbitrary trajectories and more complicated background spacetimes, such as the Kerr geometry \cite{MeentPRD94}.

Recently, a reformulation of the mode-sum regularization scheme was introduced by Wardell and Warburton~\cite{WardellPRD92} that uses \textit{tensor-harmonic regularization parameters},
\beq
\label{modesumregT}
\fF^a_\text{self}=\sum_{\ell=0}^\infty\left[\fF^{a,\ell\pm}_\text{ret}-(2\ell+1)\fF_{[-1]}^{a,\pm}-\fF_{[0]}^a\right]-D^a,
\eeq
where the \(\ell\)-modes of the retarded force are computed directly from the metric perturbation via Eq.~\eqref{fullforce},
\beq
\label{forcelmodes}
\fF^{a,\ell\pm}_\text{ret}=\left.\lim_{r\rightarrow r_0^\pm}\sum_{m=-\ell}^{\ell}\fF^{a,\ell m}[h^{\text{ret}}]\right\rvert_{\subalign{\theta&=\pi/2\\\phi&=0}},
\eeq
and the tensor-harmonic regularization parameters \(\fF_{[-1]}^{a,\pm}\), \(\fF_{[0]}^a\), and \(D^a=0\) are found by decomposing a local expansion of the Detweiler-Whiting singular field into the tensor-harmonic basis, as we will outline in Secs.~\ref{singfieldconst} and \ref{THRP}. This construction eliminates the need for re-projection onto a scalar harmonic basis and reduces the overall number of computed \(\ell\)-modes necessary to compute the regularized self-force.

\section{Regularization \label{regularization}}
The approach to self-force regularization outlined in Sec.~\ref{regscheme} was derived and implemented in the Lorenz gauge~\cite{BarackCQG26}. One might ask whether the same approach to regularization applies to other gauges, such as the RW and EZ gauges. This question was investigated by Pound, Merlin, and Barack (PMB) \cite{PoundPRD89} specifically for the radiation gauge, but their findings are equally applicable here. Under a change of gauge, \(x_\text{new}^a=x_\text{old}^a+\xi^a\), generated by a gauge vector \(\xi^a\), the metric perturbation transforms as,
\beq
\label{genericgaugetransform}
h_{ab}^\text{new}=h_{ab}^\text{old}-\pounds_\xi g_{ab}.
\eeq
Such a transformation induces a change in the self-force \cite{BarackPRD64},
\beq
\fF^a_\text{self,new}=\fF^a_\text{self,old}-\delta\fF^a_\text{self},
\eeq
with
\beq
\label{forcegaugeT}
\delta\fF^a_\text{self}=-\mu\left[(g^{ab}+u^au^b)\ddot{\xi}_b+R^a{}_{cbd}u^c\xi^bu^d\right],
\eeq
where an overdot denotes a derivative with respect to the proper time \(\tau\) of the particle's background worldline. PMB introduce a broad class of gauges under which the asymptotic matching scheme of Gralla and Wald \cite{GrallaCQG25} remains valid. This gauge class is named the \textit{sufficiently regular} gauge class. For a particular local gauge transformation away from the Lorenz gauge to remain sufficiently regular, the components of the gauge vector \(\xi^a\) must satisfy specific conditions \cite{PoundPRD89}:
\vspace{.5em}
\begin{itemize}
\setlength{\itemsep}{1em}
\item[(SR1)] \(\xi_\tau=f_1(\tau)\ln s + o(\ln s)\),
\item[(SR2)] \(\xi_i=f_2(\tau,n^i)+o(1)\),
\item[(SR3)] \(\tau\) derivatives do not increase the degree of singularity,
\item[(SR4)] spatial derivatives increase the degree of singularity by at most one order of \(s\).
\end{itemize}
Here, \(s\)  is the spatial geodesic distance away from the worldline and \(n^i\) is a spatial unit vector, expressed in local Fermi-like coordinates. For a calculation performed at first-order in the mass-ratio, \(f_1\) and \(f_2\) must be \(C^1\) almost everywhere. We demonstrate in App.~\ref{LorenzToEZ} that the local gauge transformation between the Lorenz and EZ gauges is not sufficiently regular, which motivates the adjusted approach to regularization used in this paper.

\subsection{Locally Lorenz Gauges}

To address gauge transformations away from the Lorenz gauge which are \textit{not} sufficiently regular, PMB propose the ``Locally Lorenz'' gauge (LL) regularization scheme. Beginning in the Lorenz gauge, the local metric perturbation reads~\cite{PoissonLR},
\beq
\label{lorenzsing}h_{ab}^\tL=\frac{2\mu}{s}(g_{ab}+2\tilde{u}_a\tilde{u}_b)+O(1),
\eeq
where terms of \(O(1)\) are at most bounded but discontinuous on the worldline. PMB define a gauge to be LL if it satisfies two properties: (\textit{i}) the LL metric perturbation must have an identical leading-order singular structure as the Lorenz gauge,
\beq
h_{ab}^\tLL=\frac{2\mu}{s}(g_{ab}+2\tilde{u}_a\tilde{u}_b)+o(s^{-1}),
\eeq
where terms of \(o(s^{-1})\) are not as strongly divergent as \(s^{-1}\) on the worldline, and (\textit{ii}) the Lorenz and LL gauges differ locally by at most a continuous gauge vector,~\(\xi^a_\tC\),
\beq
\label{hLLdef}
h^\tLL_{ab}=h^\tL_{ab}-\pounds_{\xi^\tC}g_{ab}.
\eeq
With these conditions in place, the two metric perturbations fall within the same class of gauges introduced by Barack and Ori \cite{BarackPRD64}, meaning that the self-forces in each gauge are related via Eq.~\eqref{forcegaugeT}.

\subsection{Regularization in the RW and EZ Gauges}
We now outline how we perform regularization in the RW/EZ gauges, motivated by the LL-gauge regularization procedure and the work of Nakano \textit{et al.} \cite{NakanoPRD68}. To start, a gauge transformation is performed locally to bring the retarded Lorenz gauge metric perturbation into the RW/EZ gauges,
 \begin{align}
 \label{retGT}
 h_{ab}^\text{RW}&=h_{ab}^\text{L}-\pounds_{\xi^\text{RW}} g_{ab}.
 \end{align}
 We perform an identical gauge transformation to a local expansion of the Detweiler-Whiting singular field \(h_{ab}^\text{L,S}\) in the Lorenz gauge,
 \beq
 \label{singularGT}
  h_{ab}^\text{RW,S}=h_{ab}^\text{L,S}-\pounds_{\xi^\text{RW,S}} g_{ab},
  \eeq
and define the difference of the two gauge vectors to be,
 \beq
 \label{SRgaugevec}
 \xi_a^\text{RW,C}\equiv \xi_a^\text{RW}-\xi_a^\text{RW,S}.
 \eeq
Assuming that \(h_{ab}^\text{L,S}\) is known to high-enough order in a series expansion \cite{HeffernanPRD86} when constructing \(\xi^\text{RW,S}\), then the remainder \(\xi^\text{RW,C}\) will be at least continuous. What exactly constitutes a ``high enough'' order is outlined in Sec.~\ref{singfieldconst}.

Using the continuous gauge vector \(\xi^\text{RW,C}_a\), we now define the LL metric perturbation from Eq.~\eqref{hLLdef} associated with the RW gauge transformation to be,
\beq
\label{hLLfromRW}
h_{ab}^\text{LL}=h_{ab}^\text{L}-\liexir.
\eeq
It must be emphasized that the LL metric perturbation in Eq.~\eqref{hLLfromRW} is not unique, as it depends on the final gauge choice enforced in Eqs.~\eqref{retGT} and \eqref{singularGT}; in general, it will differ when transforming to the RW gauge compared to the EZ gauge. Additionally, any continuous term in \(\xi_a^\text{RW,C}\) may be equally attributed to \(\xi_a^\text{RW,S}\), changing \(h^{\tLL}_{ab}\) but remaining in the Barack-Ori class. It is therefore vital that the gauge vectors \(\xi_a^\text{RW}\) and \(\xi_a^\text{RW,S}\) be specified exactly, so that we may identify \(\xi^{\text{RW,C}}_a\) in Eq.~\eqref{SRgaugevec} precisely and specify the exact LL gauge in which the regularization is performed.

To demonstrate how these gauge transformations produce an LL metric perturbation in the regularization procedure, we consider the regularization of a linear functional constructed from the metric perturbation and its derivatives, \(\mathcal{I}[h](x),\) evaluated at the spacetime event \(x\). This quantity \(\mathcal{I}\) may stand for the force in Eq.~\eqref{fullforce} or any number of gauge-invariant quantities commonly computed in the self-force literature (see e.g. Shah and Pound~\cite{ShahPRD91} for examples of these gauge invariants). We then write schematically \cite{PoundComm},
\pagebreak
 \begin{align}
\nonumber \mathcal{I}[h^\text{LL,R}](x_0)&=\lim_{x\rightarrow x_0}\mathcal{I}[h^\text{LL}-h^\text{L,S}]\\
 \nonumber &=\lim_{x\rightarrow x_0}\left\{\mathcal{I}[h^{\text{L}}-\liexir](x)-\mathcal{I}[h^{\text{L},\text{S}}]\right\}\\
 \nonumber&=\lim_{x\rightarrow x_0}\left\{\mathcal{I}[h^{\text{L}}](x)- \mathcal{I}[\pounds_{\xi^\text{RW}} g_{ab}](x)\right.\\
 \nonumber&\left.\qquad\qquad-\mathcal{I}[h^{\text{L},\text{S}}]+\mathcal{I}[\liexis](x)\right\}\\
 \label{GItransform}&=\lim_{x\rightarrow x_0}\mathcal{I}[h^{\text{RW}}-h^{\text{RW},\text{S}}](x).
 \end{align}
%
In general, the gauge term relating \(h^\text{LL}_{ab}\) and \(h^\text{L}_{ab}\) may not be dropped, and we may express the difference between the LL and Lorenz gauge quantities,
\beq
\label{forcereg}
\mathcal{I}[h^\text{LL,R}](x_0)=\mathcal{I}[h^\text{L,R}](x_0)-\mathcal{I}[\liexir](x_0).
\eeq

 The practical regularization in our work is performed by subtracting tensor-harmonic regularization terms mode-by-mode, as was done by Wardell and Warburton \cite{WardellPRD92} for Lorenz gauge regularization. For a functional of the metric perturbation, the regularization of the retarded RW gauge modes is written,
 \begin{align}
\nonumber\mathcal{I}[h^{\text{LL},\text{R}}](x_0)&=\sum_\ell\left\{\mathcal{I}^\ell[h^\text{RW}](x_0)-\mathcal{I}^\ell[h^{\text{L},\text{S}}](x_0)\right.\\
\label{modesumI}&\qquad\left.+\mathcal{I}^\ell[\liexis](x_0)\right\},
 \end{align}
where \(\mathcal{I}\) is decomposed into a tensor-harmonic basis and summed over the azimuthal index~\(m\), \ala Eq.~\eqref{forcelmodes}.
We assume that the individual \(\ell\)-modes of \(\mathcal{I}\) are continuous at the particle, and that the gauge vector in Eq.~\eqref{modesumI} is constructed solely from a local expansion of the Lorenz gauge Detweiler-Whiting singular field mode-by-mode, \(\xi^{\text{RW,S},\ell m}_a=\xi^{\text{RW},\ell m}_a[h^{\text{L},\text{S}}]\). The gauge transformation from any gauge to the RW and EZ gauges is unique in the mode-decomposition for \(\ell\ge2\), and we further outline in Sec.~\ref{l01retsol} the specific gauge choice made for \(\ell=0,1\).

Finally, we outline the regularization specifically of the self-force. Here, the \(\ell\)-modes of the retarded force contain jump discontinuities when evaluated at the particle, and the mode-sum formula is adjusted to handle these discontinuities and include the additional gauge term,
\begin{align}
\nonumber \fF_a[h^\text{LL,R}](x_0)&=\sum_\ell\left\{\fF_a^{\ell,\pm}[h^\text{RW}](x_0)-\fF_a^{\ell,\pm}[h^{\text{L},\text{S}}](x_0)\right.\\
\label{forcemodereg}&\qquad\left.+\fF_a^{\ell,\pm}[\liexis](x_0)\right\}.
\end{align}

We note that this method of self-force regularization is similar to the work of Nakano, Sago, and Sasaki \cite{NakanoPRD68}, who introduce a regularization scheme for the RW gauge analytically at 1PN based on gauge-transforming the Lorenz gauge singular field as in Eq.~\eqref{singularGT}. The methods differ in the choice of monopole and dipole gauges used in the calculation, as outlined in Sec.~\ref{l01retsol}. In addition, no post-Newtonian expansions are undertaken in our work.

\section{Retarded Solution \label{retsol}}

We now review the method used to integrate the EFEs and reconstruct the tensor-harmonic modes of the retarded metric perturbation in the EZ and RW gauges through use of \textit{master functions}, originally introduced to the study of black hole perturbation theory by Regge and Wheeler~\cite{ReggePHR108} and Zerilli~\cite{ZerilliPRD2}. We begin by introducing a tensor-harmonic basis used to decompose the metric perturbation. From the tensor-harmonic components of the metric perturbation, we construct six gauge-invariant fields used to construct the two master functions utilized in this work.

\subsection{Tensor Harmonic Decomposition}

Using the A--K framework introduced in \cite{TCW}, we take advantage of the spherical symmetry present in the Schwarzschild spacetime to decompose the metric perturbation into a basis of tensor harmonics,
\begin{equation}
h_{ab}(t,r,\theta,\phi)=\sum_{\ell=0}^\infty\sum_{m=-\ell}^\ell h_{ab}^{\ell m}(t,r,\theta,\phi),
\end{equation}
with
\begin{align}
\label{eq:hina-k}
\nonumber h_{ab}^{\ell m}(t,r,\theta,\phi)&=\mathrm{A}\,v_av_bY_{\ell m}+2\,\mathrm{B}\,v^\wg_{(a}Y^{E,\ell m}_{b)}+2\,\mathrm{C}\,v^\wg_{(a}Y^{B,\ell m}_{b)}\\
\nonumber&\quad+2\,\mathrm{D}\,v^\wg_{(a}Y^{R,\ell m}_{b)}+\mathrm{E}\,T^{T0,\ell m}_{ab}+\mathrm{F}\,T^{E2,\ell m}_{ab}\\
\nonumber&\quad+\mathrm{G}\,T^{B2,\ell m}_{ab}+2\,\mathrm{H}\,T^{E1,\ell m}_{ab}+2\,\mathrm{J}\,T^{B1,\ell m}_{ab}\\
&\quad+\mathrm{K}\,T^{L0,\ell m}_{ab},
\end{align}
where the 10 complex scalar functions A--K have had their arguments and indices suppressed for simplicity, e.g., \(\tA=\tA^{\ell m}(t,r)\). The vector and tensor harmonics are listed in App.~\ref{tensorbasis}, and the vector fields \(v_a\) and \(n_a\) are written in Schwarzschild coordinates as,
\[
v_a=(-1,0,0,0),\;\;\;n_a=(0,1,0,0).
\]
 The projection of the stress-energy, Eq.~\eqref{stressenergy}, onto the tensor-harmonic basis used in Eq.~\eqref{eq:hina-k} is straightforward, given the delta functions in the source, e.g.,
\begin{align}
\nonumber T^{\ell m}_\tA(t,r)&=f^2\int v^av^bT_{ab}Y^{*}_{\ell m}\,\mathrm{d}\Omega,\\
\nonumber&=\mu\frac{f_0\mathcal{E}}{r_0^2}\delta(r-r_0)\int Y_{\ell m}^*\delta(\theta-\pi/2)\delta(\varphi-\Omega t)\,\mathrm{d}\Omega,\\
\label{stressA}&=\mu\frac{f_0\mathcal{E}}{r_0^2}Y_{\ell m}^*(\pi/2,0)\delta(r-r_0)e^{-im\Omega t}.
\end{align}
Instead of $T^{\ell m}_\tA$ appearing explicitly, we will typically represent the occurrence of source terms by projections of the linearised Einstein operator, since by Eq.~\eqref{linearizedEFE} we have
\beq
\label{sourceprojections}
E_\tA=-16\pi T_\tA.
\eeq
All source terms relevant for circular orbits are listed in App.~\ref{sourceinAK}.

When focusing specifically on circular, equatorial orbits, the form of the source terms in Eq.~\eqref{stressA} motivates a further refinement to the Ansatz of the metric perturbtaion given in Eq.~\eqref{eq:hina-k}, whereby each scalar function A--K is written as a separable function of \(t\) and \(r\), with time-dependence of the form,
\begin{equation}
\label{trseparable}\tA^{\ell m}(t,r)=\widehat{\tA}^{\ell m}(r)\,e^{-i\omega_{ m}t}.
\end{equation}
The allowable frequencies for the metric perturbation are fixed by the source terms and are multiples of the orbital frequency,
\begin{equation}
\label{omegadefined}
\omega_m=m\Omega.
\end{equation}
This time-dependence for circular orbits is equivalent to working in the frequency domain with Fourier coefficients \cite{NakanoPRD68},
\beq
\tA^{\ell m}(t,r)=\frac1{2\pi}\int_{-\infty}^{\infty}\tA^{\ell m}(\omega,r)e^{-i\omega t}\,\rmd \omega,
\eeq
with \(\tA^{\ell m}(\omega,r)=\widehat{\tA}^{\ell m}(r)\delta(\omega-\omega_m)\).

Finally, with the introduction of the metric perturbation, certain symmetries present in the background Schwarzschild spacetime no longer exist in the physical spacetime. In particular, the vectors \((\partial_t)^a\) and \((\partial_\phi)^a\) are no longer Killing in the physical spacetime \(g_{ab}^\text{phys}\), yet a Killing vector does exist as a combination of the two: the helical Killing vector (HKV) \(k^a=(\partial_t)^a+\Omega(\partial_\phi)^a\). The physical spacetime obeys the helical symmetry \(\pounds_kg^\text{phys}=O(\mu^2/M^2)\) \cite{DetweilerPRD77}, and this symmetry exists for any reasonable choice of gauge as a consequence of the time-dependence present in Eq.~\eqref{trseparable} and the mode decomposition of the metric perturbation, Eq.~\eqref{eq:hina-k}. While we will utilize the time-dependence of Eq.~\eqref{trseparable} in this work for circular orbits, the expressions in the remainder of Sec~\ref{retsol} hold for metric perturbations with arbitrary time-dependence.


\subsection{Gauge Invariants \label{GIs}}
The procedure of metric reconstruction is based on the construction of six gauge-invariant fields introduced in \cite{TCW}; we review this construction here. We begin with the metric perturbation in Eq.~\eqref{eq:hina-k} written in an arbitrary ``old'' gauge, and write it in a ``new'' gauge by introducing a gauge vector \(\xi^a\). The transformation occurs to first-order in the mass-ratio as,
\beq
h_{ab}^\text{new}=h_{ab}^\text{old}-\pounds_\xi g_{ab}+O(\mu^2/M^2).
\eeq
The gauge vector \(\xi^a\) is decomposed into tensor-harmonic modes,
\begin{equation}
\label{gaugevec}\xi_a^{\ell m}=\tP\, v_aY_{\ell m}+\tR\, n_aY_{\ell m}+\tS\, Y^{E,\ell m}_a+\tQ\,Y_a^{B,\ell m},
\end{equation}
with complex scalar functions P, R, and S for the even-parity components of the gauge vector, and Q for the odd-parity component. The action of the gauge vector on the metric perturbation induces the following changes to the metric components:
\begin{align}
\label{delA}\Delta\tA&=-2\partial_t\tP-\frac{2Mf}{r^2}\tR,\\
\label{delB}\Delta\tB&=\frac{1}{r}\tP-\partial_t\tS,\\
\label{delC}\Delta\tC&=-\partial_t\tQ,\\
\label{delD}\Delta\tD&=\partial_r\tP-\frac{2M}{r^2f}\tP-\partial_t\tR,\\
\label{delE}\Delta\tE&=\frac{2f}{r}\tR-\frac{\lambda+2}{r}\tS,\\
\label{delF}\Delta\tF&=\frac{2}{r}\tS,\\
\label{delG}\Delta\tG&=\frac2r\tQ,\\
\label{delH}\Delta\tH&=\frac{1}{r}\tR+\partial_r\tS-\frac{1}{r}\tS,\\
\label{delJ}\Delta\tJ&=\partial_r\tQ-\frac1r\tQ,\\
\label{delK}\Delta\tK&=2\partial_r\tR+\frac{2M}{r^2f}\tR,
\end{align}
where we write e.g., \(\tA_\text{new}=\tA_\text{old}-\Delta\tA\), and introduce \(\lambda=(\ell-1)(\ell+2)\). For \(\ell\ge2\), we may enforce the gauge choice known as the Regge-Wheeler (RW) gauge, introduced by Regge and Wheeler~\cite{ReggePHR108}, by eliminating \(\tB_\text{new}=\tF_\text{new}=\tH_\text{new}=0\) through convenient choices of $\tP$, $\tS$, and $\tR$, in Eqs.~\eqref{delB}, \eqref{delF} and \eqref{delH}, and using Eq.~\eqref{delG} to eliminate \(\tG_\text{new}\). Alternatively, using Eq.~\eqref{delE} instead of Eq.~\eqref{delH}, we may set \(\tB_\text{new}=\tE_\text{new}=\tF_\text{new}=0\), which defines the EZ gauge. Specifically for the low modes \(\ell<2\), certain equations above vanish identically and another gauge choice is made that we discuss in Sec.~\ref{l01retsol}.

By combining various A--K terms and their derivatives, one may construct quantities which are unchanged under the action of the gauge vector in Eqs.~\eqref{delA}-\eqref{delK}, making them gauge-invariant:
\begin{align}
\label{alpha}\alpha&=\tJ-\frac r2\partial_r\tG,\\
\label{beta}\beta&=-\tC-\frac r2\partial_t\tG,\\
\label{chi}\chi&=\tH-\frac{1}{2f}\tE-\frac{\lambda+2}{4f}\tF-\frac{r}{2}\partial_r\tF,\\
\nonumber\psi&=\frac12\tK-\frac{r-3M}{2rf^2}\tE-\frac{r}{2f}\partial_r\tE\\
\label{psi}&\qquad-\frac{(\lambda+2)(r-3M)}{4rf^2}\tF-\frac{r(\lambda+2)}{4f}\partial_r\tF,\\
\nonumber\delta&=\tD+\frac{r}{2f}\partial_t\tE-\frac{r-4M}{rf}\tB-r\partial_r\tB\\
\label{delta}&\qquad-\frac{r^2}{2}\partial_t\partial_r\tF+\frac{r(\lambda+2)-4(r-3M)}{4f}\partial_t\tF,\\
\label{epsilon}\epsilon&=-\frac12\tA-\frac{M}{2r}\tE-r\partial_t\tB-\frac{M(\lambda+2)}{4r}\tF-\frac{r^2}{2}\partial_t^2\tF.
\end{align}

Two additional gauge invariants of interest to this work appear as combinations of certain gauge invariants above, one for each parity,
\begin{align}
\label{psiW}\Psi_\tW&=r^2\partial_t\alpha-r^2\partial_r\beta+r\beta,\\
\label{psiZ}\Psi_\tZ&=\frac{rf}{\kappa}\left[2rf\psi-r(\lambda+2)\chi\right],
\end{align}
with \(\kappa=6M+\lambda r\). These two quantities both satisfy a 1+1D wave equation in Schwarzschild time and the tortoise radial coordinate \(r_*=r+2M\log(r/2M-1)\),
\begin{equation}
\label{waveeq}
\left[-\partial_t^2+\partial_{r_*}^2-V_{\tW/\tZ}(r)\right]\Psi_{\tW/\tZ}=S_{\tW/\tZ},
\end{equation}
with potentials,
\begin{align}
\label{oddpot}V_\tW(r)&=\frac{f}{r^2}\left[\lambda+2-\frac{6M}{r}\right],\\
\label{evenpot}V_\tZ(r)&=\frac{f}{r^2}\left[\frac{\lambda^2(\lambda+2)r^3+6M(\kappa\lambda r+12M^2)}{r\kappa^2}\right].
\end{align}
We remark on the similarities between the two potentials by taking the difference,
\begin{align}
\nonumber \Delta V&=V_\tW-V_\tZ\\
&=\frac{24Mf}{r^2\kappa}\left[\left(1-\frac{3M}{r}\right)+\frac{3Mf}{\kappa}\right].
\end{align}
This difference vanishes at both the horizon and spatial infinity, and also very near but outside the light ring at \(r=3M\). It further vanishes in the limit that \(\ell\) grows to infinity.

The sources \(S_{\tW/\tZ}\) are listed in Eqs.~\eqref{sourceodd} and \eqref{sourceeven}. From Eq.~\eqref{waveeq} and the form of the potentials in Eqs.~\eqref{oddpot}-\eqref{evenpot}, it is clear that the gauge invariants \(\Psi_{\tW}\) and \(\Psi_\tZ\) are \textit{master functions} akin to those of Regge-Wheeler and Zerilli, respectively \cite{NagarCQG22}. These master functions express the two dynamical degrees of freedom in the Einstein field equations. Furthermore, it is possible to recover the gauge invariants in Eqs.~\eqref{alpha}-\eqref{epsilon} solely from the master functions, along with source terms:
\begin{align}
\label{alphafromW}\alpha&=-\frac{1}{\lambda r f}\left[\partial_t\Psi_\tW+r^2fE_\tJ\right],\\
\beta&=-\frac{1}{\lambda r}\left[f\Psi_\tW+rf\partial_r\Psi_\tW-r^3E_\tC\right],\\
\nonumber\chi&=\frac{-1}{(\lambda+2)\kappa r^2f}\left[\left\{\lambda(\lambda+2)r^2+6M(\kappa-2M)\right\}\Psi_\tZ\right.\\
&\qquad\qquad\qquad\left.+2\kappa r^2f\partial_r\Psi_\tZ+r^5E_\tA\right],\\
\nonumber\psi&=\frac{-1}{2r^2f^2\kappa}\left[2(r^2\lambda -3rM\lambda-6M^2)\Psi_\tZ\right.\\
&\qquad\qquad\qquad\left.+2r^2f\kappa\partial_r\Psi_\tZ+r^5E_\tA\right],\\
\delta&=\frac{r}{\lambda+2}\left[4f\partial_t\psi-(\lambda+2)\partial_t\chi-rE_\tD\right],\\
\label{epsilonfromZ}\epsilon&=\frac{f}{2}\left[2\chi+2rf\partial_r\chi-2f\psi+r^2E_\tF\right],
\end{align}
Thus, solving the EFEs at first order in the mass ratio has been reduced to integrating Eq.~\eqref{waveeq} for \(\Psi_{\tW/\tZ}\), up to considerations of gauge and the low modes \(\ell<2\).


\subsection{Numerical Integration}
The literature is rich with examples of numerical solutions for a point-particle source in a bound orbit about a Schwarzschild black hole, both in the time domain \cite{BarackPRD62,BarackPRD66,MartelPRD69,SopuertaPRD73,BarackPRD75} and in the frequency domain \cite{CutlerPRD47,CutlerPRD50,FujitaPTP112,AkcayPRD88,BarackPRD78}. The numerical techniques used in our work to solve the frequency-domain representation of Eq.~\eqref{waveeq},
\beq
\label{waveeqfreq}
\left[\partial_{r_*}^2+\omega_m^2-V_{\tW/\tZ}(r)\right]\widehat{\Psi}_{\tW/\tZ}=\widehat{S}_{\tW/\tZ},
\eeq
align closely with the solution method outlined by Hopper and Evans~\cite{HopperPRD82}, but simplified for the case of circular orbits. The numerical integration of Eq.~\eqref{waveeqfreq} is performed in \textsc{Mathematica}~\cite{Mathematica} to take advantage of \textsc{Mathematica}'s arbitrary precision framework. We choose to work with a global minimum precision of 32 digits, which is responsible for the ultimate numerical accuracy of the retarded field spherical harmonic modes shown later in this work.


\subsection{Metric Reconstruction}
The gauge invariants in Eqs.~\eqref{alpha}-\eqref{epsilon} may be constructed from the tensor modes of the metric perturbation \textit{in any gauge}, but play a special role in metric reconstruction specifically in the EZ gauge. When the EZ gauge conditions are enforced, Eqs.~\eqref{alpha}-\eqref{epsilon} reduce to expressions which are trivial to invert for the metric components,
\begin{align}
\label{Aez}\tA^\text{EZ}&=-2\epsilon,\\
\tC^\text{EZ}&=-\beta,\\
\tD^\text{EZ}&=\delta,\\
\tH^\text{EZ}&=\chi,\\
\tJ^\text{EZ}&=\alpha,\\
\label{Kez}\tK^\text{EZ}&=2\psi,
\end{align}
with all other components vanishing. Should one choose to work in the RW gauge instead, the non-zero metric components become,
\begin{align}
\label{Arw}\tA^\text{RW}&=-2\epsilon+\frac{2Mf}{r}\chi,\\
\tC^\text{RW}&=-\beta,\\
\tD^\text{RW}&=\delta + r\partial_t\chi,\\
\tE^\text{RW}&=-2f\chi,\\
\tJ^\text{RW}&=\alpha,\\
\label{Krw}\tK^\text{RW}&=2\psi-\frac{2(r-M)}{rf}\chi-2r\partial_r\chi.
\end{align}
The full (\(\ell\ge2\)) metric perturbation in either the EZ or RW gauge is recovered by substituting the expressions for A--K into Eq.~\eqref{eq:hina-k} after solving for the gauge-invariants via \(\Psi_{\tW/\tZ}\) in Eqs.~\eqref{alphafromW}-\eqref{epsilonfromZ}. The specific reconstruction for \(\ell<2\) is detailed in Sec.~\ref{l01retsol}.


\section{Retarded Solution for \(\ell=0,1\) \label{l01retsol}}
For the low (\(\ell<2\)) modes, the gauge invariants constructed in Sec.~\ref{retsol} lose their invariant properties under a gauge transformation. We investigate these low-order modes by gauge-transforming the Lorenz-gauge retarded solution. We opt to use the gauge choice for both \(\ell=0\) and \(\ell=1\) introduced by Zerilli \cite{ZerilliPRD2}, as the Zerilli gauge satisfies both the RW and EZ gauge conditions. This gauge choice differs from that of Nakano \textit{et al.}~\cite{NakanoPRD68}, who opt to use the Lorenz gauge monopole (corrected by Hikida \etal \cite{HikidaCQG22}) and a different variant of the Zerilli dipole.

The cases of \(\ell=0\) and \(\ell=1\) are handled separately, and the tensor-harmonic \(\ell m\) labels for the metric perturbation are written explicitly for clarity.

\subsection{\(\ell=0\) \label{l0ret}}
We approach the construction of the Zerilli gauge monopole initially by finding the gauge transformation from the Lorenz gauge to the Zerilli gauge. This will lead directly into the construction of the singular field monopole in Sec.~\ref{singfieldconst}.

At \(\ell =0\), all vector and tensor modes of the metric perturbation vanish identically. Furthermore, all coefficients of the gauge vector Eq.~\eqref{gaugevec} are evaluated with \(\omega_0=0\), eliminating any time derivatives from Eqs.~\eqref{delA}-\eqref{delK} and yielding a static gauge transformation. The gauge vector becomes,
\begin{equation}
\label{gaugevec0}
\widehat{\xi}_a^{00}=\frac{1}{2\sqrt{\pi}}\left[\widehat{\tP}^{00}\,v_a+\widehat{\tR}^{00}\,n_a\right],
\end{equation}
and induces the following changes to the metric perturbation:
\begin{align}
\label{delA0}\Delta\widehat{\tA}^{00}&=-\frac{2Mf}{r^2}\widehat{\tR}^{00},\\
\label{delD0}\Delta\widehat{\tD}^{00}&=\frac{\mathrm{d}\widehat{\tP}^{00}}{\mathrm{d}r}-\frac{2M}{r^2f}\widehat{\tP}^{00},\\
\label{delE0}\Delta\widehat{\tE}^{00}&=\frac{2f}{r}\widehat{\tR}^{00},\\
\label{delK0}\Delta\widehat{\tK}^{00}&=2\frac{\mathrm{d}\widehat{\tR}^{00}}{\mathrm{d}r}+\frac{2M}{r^2f}\widehat{\tR}^{00}.
\end{align}
The Zerilli monopole gauge choice uses the two degrees of gauge freedom to set \(\tE_{\text{Z}}^{00}=\tD_{\text{Z}}^{00}=0\). Starting from the Lorenz gauge, the choice of \(\tE^{00}_{\text{Z}}=0\) algebraically determines \(\widehat{\tR}^{00}\) from Eq.~\eqref{delE0},
\begin{equation}
\label{l0gaugevecR}\widehat{\tR}^{00}=\frac{r}{2f}\widehat{\tE}^{00}_{\text{L}}.
\end{equation}
Eq.~\eqref{delD0} is then solved to set \(\widehat{\tD}^{00}_{\text{Z}}=0\):
\begin{equation}
f\frac{\mathrm{d}}{\mathrm{d}r}\left[\widehat{\tP}^{00}/f\right]=\widehat{\tD}_{\text{L}}^{00}.
\end{equation}
When integrating this equation, we find,
\begin{equation}
\widehat{\tP}^{00}(r)=f\int_{r_1}^{r}f^{-1}(r')\widehat{\tD}^{00}_{\text{L}}(r')\,\mathrm{d}r'+f\,\widehat{\zeta}^{00}.
\end{equation}
The starting value of the integration, \(r_1\), is arbitrary, and \(\widehat{\zeta}^{00}\) is an arbitrary constant. The gauge function \(\widehat{\tP}^{00}\) is not present in the metric perturbation (outside of fixing the condition \(\widehat{\tD}^{00}_{\tZ}=0\)), as \(\widehat{\tP}^{00}\) only appears in Eq.~\eqref{delD0} for static gauge transformations. Thus, the monopole contributions to the retarded field in the Zerilli gauge are,
\begin{align}
\label{A0}\widehat{\tA}_{\text{Z}}^{00}&=\widehat{\tA}_{\text{L}}^{00}+\frac{M}{r}\widehat{\tE}_{\text{L}}^{00},\\
\label{K0}\widehat{\tK}_{\text{Z}}^{00}&=\widehat{\tK}_{\text{L}}^{00}-\frac{(r-3M)}{rf^2}\widehat{\tE}_{\text{L}}^{00}-\frac{r}{f}\frac{\mathrm{d}\widehat{\tE}_{\text{L}}^{00}}{\mathrm{d}r},
\end{align}
with all other components set to zero. These remaining components of the metric perturbation are invariant under gauge transformations produced by the gauge vector in Eq.~\eqref{gaugevec0} and are unique.

The form of the Lorenz gauge monopole was determined analytically by Barack and Lousto \cite{BarackPRD72}. The inner (\(r\le r_0\)) solution is,
\begin{align}
h_{tt}^{\text{L},-}&= -\frac{AfM}{r^3}P(r),\\
h_{rr}^{\text{L},-}&=\frac{A}{r^3f}Q(r),\\
h_{\theta\theta}^{\text{L},-}&=(\sin\theta)^{-2}h_{\varphi\varphi}^{\text{L},-}=AfP(r),
\end{align}
and the outer solution (\(r\ge r_0\)) is,
\begin{widetext}
\begin{align}
\nonumber h_{tt}^{\text{L},+}&=\frac{2\mu\mathcal{E}}{3r^4r_0f_0}\{3r^3(r_0-r)+M^2(r_0^2-12Mr_0+8M^2)\\
&\qquad+(r_0-3M)[-rM(r+4M)+rP(r)f\ln f+8M^3\ln(r_0/r)]\},\\
\nonumber h_{rr}^{\text{L},+}&=-\frac{2\mu\mathcal{E}}{3r^4r_0f_0f^2}\{-r^3r_0-2Mr(r_0^2-6Mr_0-10M^2)+3M^2(r_0^2-12Mr_0+8M^2)\\
&\qquad+(r_0-3M)[5Mr^2+(r/M)Q(r)f\ln f-8M^2(2r-3M)\ln(r_0/r)]\},\\
\nonumber h_{\theta\theta}^{\text{L},+}&=(\sin\theta)^{-2}h_{\varphi\varphi}^{\text{L},+}=-\frac{2\mu\mathcal{E}}{9rr_0f_0}\{3r_0^2M-80M^2r_0+156M^3\\
&\qquad+(r_0-3M)[-3r^2-12Mr+3(r/M)P(r)f\ln f+44M^2+24M^2\ln(r_0/r)]\}.
\end{align}
\end{widetext}
The constant \(A\) and the functions \(P(r)\) and \(Q(r)\) were originally introduced by Barack and Lousto,
\begin{align}
A&=\frac{2\mu\mathcal{E}}{3Mr_0f_0}[M-(r_0-3M)\ln f_0],\\
P(r)&=r^2+2Mr+4M^2,\\
Q(r)&=r^3-Mr^2-2M^2r+12M^3,
\end{align}
with \(f_0=f(r_0)\), and are not to be confused with quantities elsewhere in this work. Before we perform the gauge transformation in Eqs.~\eqref{A0}-\eqref{K0}, it is important to realize that the Lorenz gauge monopole is not asymptotically flat (in this instance, defined as \(h_{tt}^{+}\rightarrow 0\) as \(r\rightarrow \infty\)),
\begin{equation}
h_{tt}^{\text{L}+}=-\frac{2\mu\mathcal{E}}{r_0f_0}(1-\frac{r_0}{r})+O(1/r^2)\;\;\;\text{as}\;r\rightarrow\infty,
\end{equation}
and so we choose to perform an additional gauge transformation to adjust this after transforming to the Zerilli gauge. The asymptotic flatness of the monopole is important for the comparison between gauge-invariants for the purposes of this work \cite{SagoPRD78}. We shall see that the gauge vector required for this transformation \textit{does not} obey the HKV symmetry, and would not be attainable via the gauge vector in Eq.~\eqref{gaugevec0}.

After constructing the metric components in the Zerilli gauge via Eqs.~\eqref{A0}-\eqref{K0} and recovering the full metric perturbation from Eq.~\eqref{eq:hina-k}, the not-asymptotically-flat (NAF) Zerilli gauge metric is,
\begin{align}
h_{tt}^{\tZ,\NAF-}&=0,\\
h_{rr}^{\tZ,\NAF-}&=0,\\
h_{tt}^{\tZ,\NAF+}&=-\frac{2\mu\mathcal{E}(r-r_0)}{rr_0f_0},\\
h_{rr}^{\tZ,\NAF+}&=\frac{2\mu\mathcal{E}}{rf}.
\end{align}
To correct \(h^{\tZ,\NAF+}_{tt}\), we introduce a gauge vector taking the form of a global homogeneous solution to Eq.~\eqref{delD0} which breaks the HKV symmetry, i.e., has non-vanishing time dependence, but maintains the Zerilli gauge condition,
\begin{equation}
\label{NAFgaugevec}
\xi_a^{\NAF}=\frac{\mu\mathcal{E}}{r_0f_0} tfv_a.
\end{equation}
 This gauge vector changes the \(tt\)-component of the metric via Eq.~\eqref{delA} to,
\begin{align}
h_{tt}^{\tZ,-}&=\frac{2\mu\mathcal{E}}{r_0f_0} f,\\
h_{rr}^{\tZ,-}&=0,\\
h_{tt}^{\tZ,+}&=\frac{2 \mu\mathcal{E}}{r},\\
h_{rr}^{\tZ,+}&=\frac{2\mu\mathcal{E}}{rf},
\end{align}
and the perturbation now vanishes at both the horizon and spatial infinity. We notice that, while \(h_{tt}^{\tZ}\) is continuous across the particle's orbit, a jump-discontinuity has been introduced to \(h_{rr}^{\tZ}\) that was not present in the Lorenz gauge.

\subsection{\(\ell=1\) Odd-Parity \label{l1oddret}}
For \(\ell=1\), the only non-zero odd-parity contribution to the metric perturbation arises from \(m=0\). Furthermore, the spin-2 contribution to the metric perturbation, \(\tG^{10}\), vanishes identically, and Zerilli chooses to use the one degree of gauge freedom, \(\tQ^{10}\), to eliminate \(\tJ_{\text{Z}}^{10}=0\). This gauge choice is identical to the odd-parity dipole gauge used in Lorenz gauge calculations \cite{DetweilerPRD69}. Its derivation may be found in the literature, for example from~\cite{TCW}, and the analytic solution is given by,
\begin{align}
h^{\text{Z},-}_{t\phi}&=-2\mu\mathcal{L}\sin^2\theta\frac{r^2}{r_0^3},\\
h^{\text{Z},+}_{t\phi}&=-2\mu\mathcal{L}\sin^2\theta\frac{1}{r}.
\end{align}

\subsection{\(\ell=1\) Even-Parity \label{l1evenret}}
Restricting to \(\ell=1\) even-parity, the metric perturbation vanishes for \(m=0\), so only the values \(m=\pm1\) need be considered. Unlike for \(\ell=0\) and \(\ell=1\) odd-parity, there are no known analytic solutions for the even-parity dipole in the Lorenz gauge. Despite this lack of analytic solution, we work through the gauge transformation required to bring the Lorenz gauge solution to the Zerilli gauge, as this transformation will be required to construct the even-parity dipole singular field in Sec.~\ref{singfieldconst}. Analytic solutions to the even-parity dipole do exist in the Zerilli gauge, which we list at the end of this section.

The changes to the metric perturbation under a gauge transformation reduce for \(\ell=1\) even parity to,
\begin{align}
\label{delA1}\Delta\widehat{\tA}^{1m}&=2i\omega_m\widehat{\tP}^{1m}-\frac{2Mf}{r^2}\widehat{\tR}^{1m},\\
\label{delB1}\Delta\widehat{\tB}^{1m}&=\frac{1}{r}\widehat{\tP}^{1m}+i\omega_m\widehat{\tS}^{1m},\\
\Delta\widehat{\tD}^{1m}&=\frac{\mathrm{d}\widehat{\tP}^{1m}}{\mathrm{d}r}-\frac{2M}{r^2f}\widehat{\tP}^{1m}+i\omega_m\widehat{\tR}^{1m},\\
\label{delE1}\Delta\widehat{\tE}^{1m}&=\frac{2f}{r}\widehat{\tR}^{1m}-\frac{2}{r}\widehat{\tS}^{1m},\\
\label{delH1}\Delta\widehat{\tH}^{1m}&=\frac{1}{r}\widehat{\tR}^{1m}+\frac{\mathrm{d}\widehat{\tS}^{1m}}{\mathrm{d}r}-\frac{1}{r}\widehat{\tS}^{1m},\\
\Delta\widehat{\tK}^{1m}&=2\frac{\mathrm{d}\widehat{\tR}^{1m}}{\mathrm{d}r}+\frac{2M}{r^2f}\widehat{\tR}^{1m}.
\end{align}
Here, \(\widehat{\tF}^{1 m}=0\) identically but we still have the full even-parity gauge freedom. The Zerilli dipole gauge is determined by setting \(\widehat{\tB}^{1m}_\text{Z}=\widehat{\tE}^{1m}_\text{Z}=\widehat{\tH}^{1m}_\text{Z}=0\), and the gauge vector for this choice is calculated in two steps, where first \(\widehat{\tP}^{1m}\) and \(\widehat{\tR}^{1m}\) are found algebraically via Eqs.~\eqref{delB1} and \eqref{delE1} while leaving \(\widehat{\tS}^{1m}\) free,
\begin{align}
\widehat{\tP}^{1m}_\tZ&=r\left(\widehat{\tB}^{1m}_\text{L}-i\omega_m\widehat{\tS}^{1m}_\tZ\right),\\
\label{Rl1}\widehat{\tR}^{1m}_\tZ&=\frac{r}{2f}\left(\widehat{\tE}^{1m}_\text{L}+\frac2r\widehat{\tS}^{1m}_\tZ\right),
\end{align}
which, when substituted into Eq.~\eqref{delH1}, yield a first-order ODE for \(\widehat{\tS}^{1m}_\tZ\):
\begin{equation}
\frac{\mathrm{d}}{\mathrm{d}r}\left(f\,\widehat{\tS}_\tZ^{1m}\right)=f\,\widehat{\tH}^{1m}_\text{L}-\frac{1}{2}\widehat{\tE}^{1m}_\text{L}.
\end{equation}
The solution may be found by integration,
\beq
\widehat{\tS}^{1m}_\tZ=\widehat{\tS}^{1m}_\text{part}+f^{-1}\widehat{\zeta}^{1m},
\eeq
where \(\widehat{\zeta}^{1m}\) is a constant and we have written,
\beq
\label{Sdefined}
\widehat{\tS}^{1m}_\text{part}(r)\equiv f^{-1}\int_{r_0}^r\left(f(r')\widehat{\tH}^{1m}_{\text{L}}(r')-\frac{1}{2}\widehat{\tE}^{1m}_{\text{L}}(r')\right)\,\mathrm{d}r'.
\eeq

When transforming from the Lorenz gauge in the region around the particle's orbit, \(\widehat{\tH}^{1m}_\text{L}(r)\) and \(\widehat{\tE}^{1m}_\text{L}(r)\) are both bounded, \(C^0\) functions of \(r\), and thus \(\widehat{\tS}^{1m}_\text{part}(r)\) is a \(C^1\) function over the same interval. The lower bound for the integral in \(\widehat{\tS}^{1m}_\text{part}\) is arbitrary and set to the orbital radius for convenience, such that \(\widehat{\tS}_\text{part}^{1m}\) vanishes at the particle (but note that its radial derivative does not vanish). In addition, the unknown constant \(\widehat{\zeta}^{1m}\) is arbitrary.
After the gauge transformation, the remaining non-zero components of the metric are,
\begin{widetext}
\begin{align}
\label{zerillidipoleA}\widehat{\tA}^{1m}_{\text{L}\rightarrow\tZ}&=\widehat{\tA}^{1m}_\text{L}-2i\omega_mr\widehat{\tB}^{1m}_\text{L}+\frac{M}{r}\widehat{\tE}^{1m}_\text{L}+2r\left(\frac{M}{r^3}-\omega_m^2\right)\widehat{\tS}^{1m}_\text{part}-\Delta\widehat{\tA}^{1m}_\tZ,\\
\label{zerillidipoleD}\widehat{\tD}^{1m}_{\text{L}\rightarrow\tZ}&=\widehat{\tD}^{1m}_\text{L}-\left(\frac{r-4M}{rf}+r\frac{\mathrm{d}}{\mathrm{d}r}\right)\widehat{\tB}^{1m}_\text{L}-\frac{i\omega_mr}{2f}\widehat{\tE}^{1m}_\text{L}+i\omega_m\left(r\frac{\mathrm{d}}{\mathrm{d}r}-\frac{4M}{rf}\right)\widehat{\tS}^{1m}_\text{part}-\Delta\widehat{\tD}^{1m}_\tZ,\\
\label{zerillidipoleK}\widehat{\tK}^{1m}_{\text{L}\rightarrow\tZ}&=\widehat{\tK}^{1m}_\text{L}-\left(\frac{(r-3M)}{rf^2}+\frac{r}{f}\frac{\mathrm{d}}{\mathrm{d}r}\right)\widehat{\tE}^{1m}_\text{L}-\left(\frac{2}{f}\frac{\mathrm{d}}{\mathrm{d}r}-\frac{2M}{r^2f^2}\right)\widehat{\tS}^{1m}_\text{part}-\Delta\widehat{\tK}^{1m}_\tZ,
\end{align}
\end{widetext}
with residual gauge freedom,
\begin{align}
\nonumber\Delta\widehat{\tA}_\tZ^{1m}&=-\frac{2r}{f}\left(\frac{M}{r^3}-\omega_m^2\right)\widehat{\zeta}^{1m},\\
\label{zerillihomogeneous}\Delta\widehat{\tD}_\tZ^{1m}&=\frac{6i\omega_mM}{rf^2}\widehat{\zeta}^{1m},\\
\nonumber\Delta\widehat{\tK}_\tZ^{1m}&=-\frac{6M}{r^2f^3}\widehat{\zeta}^{1m}.
\end{align}

While it is clear that the metric perturbation in Eqs.~\eqref{zerillidipoleA}-\eqref{zerillidipoleK} is in the Zerilli gauge, the additional gauge freedom in Eqs.~\eqref{zerillihomogeneous} may be added to the metric perturbation without changing the gauge condition \(\tB^{1m}_\tZ=\tE^{1m}_\tZ=\tH^{1m}_\tZ=0\), and so the gauge choice is not uniquely fixed. We now use this freedom to recover a Zerilli gauge in which all components of the metric perturbation vanish outside the particle's orbit (\(r>r_0\)). This choice is made to ensure that the dipole is asymptotically flat.

We begin with the analytic, retarded Zerilli gauge solution given by Detweiler and Poisson \cite{DetweilerPRD69},
\begin{align}
\nonumber h^\tZ_{tt}&=\frac{2\mu r_0f_0\mathcal{E}}{r^2f}\left(1-\frac{r^3\Omega^2}{M}\right)\sin\theta\\
&\qquad\qquad\qquad\times\cos(\phi-\Omega t)\Theta(r-r_0),\\
h^\tZ_{tr}&=-\frac{6\mu r_0f_0\Omega \mathcal{E}}{rf^2}\sin\theta\sin(\phi-\Omega t)\Theta(r-r_0),\\
h^\tZ_{rr}&=\frac{6\mu r_0f_0\Omega \mathcal{E}}{r^2f^3}\sin\theta\cos(\phi-\Omega t)\Theta(r-r_0),
\end{align}
where \(\Theta(r-r_0)\) is the Heaviside step function. Transforming this solution to one which vanishes in the outer region via Eqs.~\eqref{zerillihomogeneous} and factoring out the time-dependence yields the A--K components of the metric perturbation,
\begin{align}
\label{zerilliAret}\widehat{\tA}_\tZ^{1m}&=\frac{2r}{f}\left(\frac{M}{r^3}-\omega_m^2\right)\frac{r_0^3\bar{E}^{1m}_\tA}{12M}\Theta(r_0-r),\\
\label{zerilliDret}\widehat{\tD}_\tZ^{1m}&=-\frac{6i\omega_mM}{rf^2}\frac{r_0^3\bar{E}_\tA^{1m}}{12M}\Theta(r_0-r),\\
\label{zerilliKret}\widehat{\tK}_\tZ^{1m}&=\frac{6M}{r^2f^3}\frac{r_0^3\bar{E}_\tA^{1m}}{12M}\Theta(r_0-r),
\end{align}
 where \(\bar{E}_\tA^{1m}\) is the fully-evaluated coefficient of the delta function source in Eq.~\eqref{eq:stressA}. By inspection, this solution is almost entirely pure gauge; for both \(r<r_0\) and \(r>r_0\), the form of Eqs.~\eqref{zerilliAret}-\eqref{zerilliKret} is identical to Eqs.~\eqref{zerillihomogeneous} with particular choices for \(\widehat{\zeta}^{1m}\) in each domain. Truly, it is the step function itself that makes the solution physically meaningful, as otherwise the entire metric perturbation in this sector may be set to vanish by choosing the appropriate constant in Eqs.~\eqref{zerillihomogeneous}.

 We now wish to refine the gauge transformation used to recover Eqs.~\eqref{zerillidipoleA}-\eqref{zerillidipoleK} from the Lorenz gauge to the particular Zerilli gauge used in Eqs.~\eqref{zerilliAret}-\eqref{zerilliKret}, which will exhaust all of the remaining gauge freedom generated by a gauge vector obeying the helical symmetry. Our choice is to eliminate the right-hand-sided limit of the Zerilli metric perturbation generated from the Lorenz gauge solution at the particle,
\begin{equation}
\label{limits}\lim_{r\rightarrow r_0^+}\widehat{\tA}^{1m}_{\text{L}\rightarrow\tZ}(r)=\lim_{r\rightarrow r_0^+}\widehat{\tD}^{1m}_{\text{L}\rightarrow\tZ}(r)=\lim_{r\rightarrow r_0^+}\widehat{\tK}^{1m}_{\text{L}\rightarrow\tZ}(r)=0.
\end{equation}
This gauge refinement condition may be enforced at any value of \(r>r_0\), but we choose to evaluate (the right-hand-sided limit) at \(r=r_0\), since we have constructed \(\tS_\text{part}^{1m}\) to vanish at the orbit, which greatly simplifies Eqs.~\eqref{zerillidipoleA}-\eqref{zerillidipoleK}.

\begin{widetext}
As \(\widehat{\tS}^{1m}_\text{part}(r)\) is a differentiable function, we find,
\begin{align}
\nonumber\lim_{r\rightarrow r_0}\frac{\mathrm{d}\widehat{\tS}^{1m}_\text{part}}{\mathrm{d}r}(r)&=f_0^{-1}\lim_{r\rightarrow r_0}\frac{\mathrm{d}}{\mathrm{d}r}\int_{r_0}^r\left(f(r')\widehat{\tH}^{1m}_\text{L}(r')-\frac{1}{2}\widehat{\tE}^{1m}_\text{L}(r')\right)\,\mathrm{d}r'-\frac{2M}{r_0^2f_0^2}\widehat{\tS}^{1m}_\text{part}(r_0),\\
&=\widehat{\tH}^{1m}_\text{L}(r_0)-\frac{1}{2f_0}\widehat{\tE}^{1m}_\text{L}(r_0).
\end{align}
Then, after taking the limits in Eq.~\eqref{limits},
\begin{align}
\label{Avanisheq}0&=\widehat{\tA}^{1m}_\text{L}-2i\omega_mr_0\widehat{\tB}^{1m}_\text{L}+\frac{M}{r_0}\widehat{\tE}^{1m}_\text{L},\\
\label{Dvanisheq}0&=\widehat{\tD}^{1m}_\text{L}-\frac{r_0-4M}{r_0f_0}\widehat{\tB}^{1m}_\text{L}-r_0\left(\frac{\mathrm{d}\widehat{\tB}^{1m}_\text{L}}{\mathrm{d}r}\right)_+-\frac{i\omega_mr_0}{f_0}\widehat{\tE}^{1m}_\text{L}+i\omega_mr_0\widehat{\tH}^{1m}_\text{L}-\frac{6i\omega_mM}{r_0f_0^2}\widehat{\zeta}^{1m},\\
\label{Kvanisheq}0&=\widehat{\tK}^{1m}_\text{L}+\frac{3M}{r_0f_0^2}\widehat{\tE}^{1m}_\text{L}-\frac{r_0}{f_0}\left(\frac{\mathrm{d}\widehat{\tE}^{1m}_\text{L}}{\mathrm{d}r}\right)_+-\frac{2}{f_0}\widehat{\tH}^{1m}_\text{L}+\frac{6M}{r_0^2f_0^3}\widehat{\zeta}^{1m}.
\end{align}
The validity of this choice must now be verified.

We begin by analyzing Eq.~\eqref{Avanisheq}. In the Zerilli gauge, \(\widehat{\tA}^{1m}_\tZ\) is gauge-invariant at the particle, which can be seen by substituting Eqs.~\eqref{delA1}-\eqref{delE1} into the combination of metric components found in Eq.~\eqref{Avanisheq},
\begin{equation}
\label{deltaAZ}
\Delta\widehat{\tA}-2i\omega_mr\Delta\widehat{\tB}+\frac{M}{r}\Delta\widehat{\tE}=-2r\left(\frac{M}{r^3}-\omega_m^2\right)\tS^{1m}.
\end{equation}
This combination vanishes at the particle irrespective of the choice of \(\widehat{\tS}^{1m}\), since \(\omega_m^2=\Omega^2\) for \(m=\pm1\). Thus, if \(\widehat{\tA}^{1m}_\tZ(r_0)\) vanishes in one gauge, it must vanish in all gauges related via the HKV symmetry. This result is unsurprising; \(\widehat{\tA}^{1m}_\tZ(r_0)\) is the sole contribution to the even-parity piece of the Detweiler redshift invariant~\(\bar{u}^t\) in the Zerilli gauge (where \(h_{t\phi}=h_{\phi\phi}=0\) for even-party). Since \(\widehat{\tA}^{1m}_\tZ(r_0)\) vanishes in both the left- and right-hand-sided limits in the Zerilli gauge, as shown in Eq.~\eqref{zerilliAret}, the condition Eq.~\eqref{limits} is satisfied for \(\widehat{\tA}^{1m}_{\text{L}\rightarrow\tZ}\).

To show that the remaining two limits are valid requires more work, and we must solve for \(\widehat{\zeta}^{1m}\) to satisfy the vanishing conditions. Both Eqs.~\eqref{Dvanisheq} and \eqref{Kvanisheq} provide a solution for the remaining gauge freedom and the system appears overdetermined.
We solve both equations,
\begin{align}
\label{zetaD}\widehat{\zeta}^{1m}_\text{D}&=\frac{f_0}{6\omega_mM}\left[i(r_0-4M)\widehat{\tB}^{1m}_\text{L}+ir_0^2f_0\left(\frac{\mathrm{d}\widehat{\tB}^{1m}_\text{L}}{\mathrm{d}r}\right)_+-ir_0f_0\widehat{\tD}^{1m}_\text{L}-\omega_m r_0^2\widehat{\tE}^{1m}_\text{L}+\omega_m r_0^2f_0\widehat{\tH}^{1m}_\text{L}\right],\\
\label{zetaK}\widehat{\zeta}^{1m}_\text{K}&=\frac{r_0f_0}{6M}\left[-3M\widehat{\tE}^{1m}_\text{L}+r_0^2f_0\left(\frac{\mathrm{d}\widehat{\tE}^{1m}_\text{L}}{\mathrm{d}r}\right)_++2r_0f_0\widehat{\tH}^{1m}_\text{L}-r_0f_0^2\widehat{\tK}^{1m}_\text{L}\right],
\end{align}
labeling the solution for \(\widehat{\zeta}^{1m}_\text{D/K}\) arising from each equation separately. The difference between these two constants is proportional to a source term,
\beq
\label{l1verify}\widehat{\zeta}^{1m}_{\text{K}}-\widehat{\zeta}^{1m}_{\text{D}}=\frac{r_0^3f_0^2}{12iM\omega_m} E^{1m}_\tD,
\eeq
and this source term vanishes for the circular orbits of interest in this paper, \(E^{1m}_\tD=0\). The constant may then be determined by use of either Eq.~\eqref{Dvanisheq} or \eqref{Kvanisheq}, and the gauge freedom is now entirely fixed. The vanishing right-hand side of Eq.~\eqref{l1verify} is verified numerically in Sec.~\ref{results}.
\end{widetext}


\section{Singular Field Construction \label{singfieldconst}}
In this section we construct the Detweiler-Whiting singular field in the EZ and RW gauges. We begin with a local expansion of the singular field in the Lorenz gauge. After a decomposition into tensor harmonic modes, the gauge-invariants Eqs.~\eqref{alpha}-\eqref{epsilon} are formed and used to reconstruct the singular field in both the EZ and RW gauges via Eqs.~\eqref{Aez}-\eqref{Kez} and Eqs.~\eqref{Arw}-\eqref{Krw}, respectively. We then detail the specific gauge transformation of the singular field for the low-order (\(\ell<2\)) modes.


\subsection{Local Detweiler-Whiting Singular Field}
The trace-reversed Detweiler-Whiting singular field is found in the Lorenz gauge and expanded covariantly about the worldline of the particle \cite{HeffernanPRD86},
\begin{equation}
\label{hbarS}
\bar{h}^{\text{L},\text{S}}_{ab}=4\mu g_{a}{}^{\bar{a}}g_{b}{}^{\bar{b}}\left[\frac1\varepsilon\frac{u_{\bar{a}}u_{\bar{b}}}{\bar{s}}+O(\varepsilon)\right],
\end{equation}
with \(u_{\bar{a}}\) and \(g_{\bar{a}\bar{b}}\) the particle's four-velocity and the background metric, respectively, evaluated on the worldline, \(g_{a}{}^{\bar{a}}\) the bivector of parallel transport, \(\bar{s}=(g_{\bar{a}\bar{b}}+u_{\bar{a}}u_{\bar{b}})\sigma^{\bar{a}}\sigma^{\bar{b}}\) the spatial geodesic distance away from the worldline, and \(\sigma\) the Synge world function. \(\varepsilon\) is an order-counting parameter in the expansion. (See \cite{PoissonLR} for a review of bitensors and covariant expansions of \(h^\text{S}\).) Following conventions established in the self-force literature \cite{BarackPRD61,DetweilerPRD67,HaasPRD74}, a coordinate expansion of Eq.~\eqref{hbarS} is performed in coordinates (\(\Delta t,\Delta r,\Theta,\Phi\)) about some reference Schwarzschild time \(t_0=0\), such that \(\Delta t=0\), \(\Delta r=r-r_0\), and the angles (\(\Theta,\Phi\)) are related to the background Schwarzschild angles (\(\theta,\phi\)) by the rotation,
\begin{align}
\nonumber\sin\theta\cos\phi&=\cos\Theta,\\
\nonumber\sin\theta\sin\phi&=\sin\Theta\cos\Phi,\\
\label{rotation}\cos\theta&=\sin\Theta\sin\Phi.
\end{align}
This rotation places the particle at the pole of the rotated coordinates, (\(\theta=\pi/2,\phi=0)\rightarrow(\Theta=0,\Phi \text{ arbitrary})\). In these coordinates the field has the form \cite{WardellPRD92},
\begin{equation}
\bar{h}_{ab}^{\text{L},\text{S}}=\frac 1\varepsilon \frac{c_{ab}^{(1)}}{\rho}+\varepsilon^0\left[\frac{c_{ab}^{(2)}\Delta r}{\rho}+\frac{c_{ab}^{(3)}\Delta r^3}{\rho^3}\right]+O(\varepsilon),
\end{equation}
evaluated at \(\Delta t=0\), where, for the circular orbits of interest in this paper, the coefficients \(c_{ab}^{(n)}\) are independent of \(\Delta r\) and \(\Theta\), and we have introduced \(\rho\) as the leading-order term in the coordinate expansion of \(\bar{s}\) \cite{DetweilerPRD67},
\beq
\label{definerho}
\rho^2=\frac{2\gamma r_0^2}{r_0-3M}(\nu^2+1-\cos\Theta),
\eeq
with
\beq
\label{definegamma}
\gamma=1-\frac{M}{r_0f_0}\sin^2\Phi,
\eeq
and
\beq
\label{definenu}
\nu^2=\frac{r_0-3M}{r_0^3f_0^2}\frac{\Delta r^2}{2\gamma}.
\eeq
The full coordinate expansion of \(\bar{h}^\text{L,S}_{ab}\) used for this work is quite lengthy, so we direct the reader to an online source for the expansion through \(O(\varepsilon^4)\) \cite{BarryWebsite}. We include orders up through \(O(\varepsilon^2)\), in order to capture the necessary angular derivatives required to regularize the EZ-gauge self-force.


\subsection{Tensor Harmonic Decomposition of \(h_{ab}^\text{L,S}\)\label{harmonicS}}
To find the tensor harmonic projections of the singular field, we follow the work of Wardell and Warburton \cite{WardellPRD92}, who calculate the tensor modes of the singular field in the BLS basis. We outline the relationship between the BLS basis and the A--K basis in App.~\ref{BLSbasis}. Our construction of the singular field modes is identical to \cite{WardellPRD92}.

Before we begin, it is worth recalling that in the rotated coordinates the particle is located at the pole  (\(\Theta=0,\Phi\text{ arbitrary}\)). When decomposed into tensor-harmonic \(\ell m'\)-modes in these rotated coordinates, the tensor harmonic basis vanishes at the particle for all but select values of \(m'\) (the azimuthal index number associated with \(\Phi\)), and so only these non-vanishing \(m'\) modes of the singular field are required. The required A--K terms for each \(m'\) are listed in Table~\ref{mpvalues}.

We demonstrate the process of finding the tensor-harmonic decomposition of the singular field for the A term through \(O(\Delta r)\), for simplicity. Starting with the projection,

\begin{table}
\caption{\label{mpvalues} We list the A--K components of the Lorenz gauge singular field required for this work at each \(m'\) value considered in this construction.}
\begin{ruledtabular}
\begin{tabular}{l l}
\(m'\)&Non-vanishing A--K\\
\hline \\
0&A, E, F, H, K\\
1&B, C, D\\
2&A, E, F, G, H, J, K\wa\wa\wa\wa\wa\wa\wa\wa\wa\\
\end{tabular}
\end{ruledtabular}
\end{table}
\begin{align}
\label{atermSintegrals}\tA_{\text{L},\tS}^{\ell 0'}&=f^2\int v^av^bh^{\text{L},\text{S}}_{ab}Y_{\ell 0}^*\,\rmd\Omega\\
\nonumber&=\sqrt{\frac{2\ell+1}{4\pi}}\int_0^{2\pi}\int_0^\pi h^{\text{L},\tS}_{tt}P_\ell(\cos\Theta)\sin\Theta\,\rmd\Theta\,\rmd\Phi,
\end{align}
we substitute in the coordinate expansion for \(h_{tt}^{\text{L},\tS}=\bar{h}_{tt}^{\text{L},\tS}-\frac12g_{tt}g^{cd}\bar{h}_{cd}^{\text{L},\tS}\), with the trace-reversed singular field given through \(O(\Delta r)\) by,
\begin{align}
\bar{h}_{tt}^{\text{L},\tS}&=\frac1\rho\left[\frac{4r_0^2f_0^2}{r_0(r_0-3M)}-\frac{2\Delta r}{r_0^2(r_0-3M)}\right.\\
\nonumber&\left.\times\frac{r_0^2-7Mr_0+10M^2-2r_0f_0(r_0-4M)(1-\gamma)}{\gamma}\right],\\
\bar{h}_{rr}^{\text{L},\tS}&=0,\\
\bar{h}_{\Theta\Theta}^{\text{L},\tS}&=\left[1-\frac{r_0f_0(1-\gamma)}{M}\right]^2\; \bar{h}_\text{ang}^{\text{L},\tS},\\
\bar{h}_{\Phi\Phi}^{\text{L},\tS}&=\frac{r_0f_0(1-\gamma)}{M}\left[1-\frac{r_0f_0(1-\gamma)}{M}\right]\sin^2\Theta\;\bar{h}_\text{ang}^{\text{L},\tS},
\end{align}
with
\begin{align}
\bar{h}_\text{ang}^{\text{L},\tS}&=\frac{1}{\rho}\left[\frac{4Mr_0^2}{r_0-3M}+\frac{2Mr_0\Delta r}{r_0-3M}\right.\\
\nonumber&\left.\qquad\;\;\;\;\times\frac{3r_0-7M-2r_0f_0(1-\gamma)}{r_0f_0\gamma}\right],
\end{align}
and the \(\Phi\)-dependence expressed through \(\gamma\).

The integral over \(\Theta\) is performed first. Recall from Eq.~\eqref{definerho} that \(\rho\) has \(\Theta\)-dependence. As such, the integral over \(\Theta\) becomes,
\pagebreak
\begin{align}
\label{thetaint}
\int_0^\pi&\frac{P_\ell(\cos\Theta)\sin\Theta}{\rho}\,\rmd\Theta\\
\nonumber&\sim\int_{-1}^1\frac{P_\ell(\cos\Theta)}{(\nu^2+1-\cos\Theta)^{1/2}}\,\rmd(\cos\Theta),
\end{align}
neglecting factors in \(\rho\) that do not depend on \(\Theta\). The denominator of Eq.~\eqref{thetaint} is expandable in terms of Legendre polynomials \cite{KeidlPRD82}, and for \(\nu\sim\Delta r\ll1\) but finite,
\begin{align}
\label{legendreexp}
\frac1{(\nu^2+1-\cos\Theta)^{1/2}}&\\
\nonumber=\sum_{\ell'} \left[\sqrt{2}-\right.&\left.(2\ell'+1)|\nu|+O(\nu^2)\right]P_{\ell'}(\cos\Theta).
\end{align}
 Eq.~\eqref{thetaint} is written, using Eq.~\eqref{legendreexp} and substituting \(u=\cos\Theta\), as,
 \begin{align}
\int_{-1}^1&\frac{P_\ell(u)}{(\nu^2+1-u)^{1/2}}\,\rmd u\\
\nonumber=&\sum_{\ell'}\left[\sqrt{2}-(2\ell'+1)|\nu|+O(\nu^2)\right]\int_{-1}^1P_{\ell'}(u)P_{\ell}(u)\,\rmd u\\
\nonumber =&\sum_{\ell'}\left[\sqrt{2}-(2\ell'+1)|\nu|+O(\nu^2)\right]\left(\frac{2}{2\ell'+2}\right)\delta_{\ell\ell'}\\
 \nonumber=&\left(\frac{2}{2\ell+1}\right)\left[\sqrt{2}-(2\ell+1)|\nu|+O(\nu^2)\right],
 \end{align}
where the third line follows from the orthogonality of the Legendre polynomials, and \(\delta_{\ell\ell'}\) is the Kronecker delta. This result is the integral over \(\Theta\) expanded as a power series in \(\nu\).

After integrating over \(\Theta\), we focus on the integral over \(\Phi\). All \(\Phi\)-dependence is now found in fractional or whole powers of \(\gamma\), and the integral of these terms becomes a hypergeometric function \cite{WardellPRD92},
\beq
\label{phiint}
\int_0^{2\pi}\gamma^n\,\rmd\Phi=2\pi\,{}_2F_1\left(n,\frac12,1,\frac{M}{r_0f_0}\right).
\eeq
When \(n=-1/2\), the integral is proportional to the elliptic integral of the first kind, \(\hat{\mathcal{K}}(\frac{M}{r_0f_0})\), and when \(n=1/2\) it is proportional to the elliptic integral of the second kind, \(\hat{\mathcal{E}}(\frac{M}{r_0f_0})\). All integer values of \(n\) reduce Eq.~\eqref{phiint} to a polynomial in \(\frac{M}{r_0f_0}\), and any other value of \(n\) is related to these three cases by the recursion relation for \(\mathcal{F}_p(k)\equiv{}_2F_1(p,\frac12,1,k)\),
\beq
\label{Frecursion}
\mathcal{F}_{p+1}(k)=\frac{p-1}{p(k-1)}\mathcal{F}_{p-1}(k)+\frac{1-2p+(p-\frac12)k}{p(k-1)}\mathcal{F}_p(k).
\eeq

When the dust has settled, \(A_\tS^{\ell 0}\) is given to linear order in \(\Delta r\) as,
\begin{widetext}
\begin{align}
\nonumber\tA_{\text{L},\tS}^{\ell 0'}&=\sqrt{\frac{4\pi}{(2\ell+1)(r_0-3M)}}\left[
-\frac{4(r_0-M)f_0^{1/2}\mathcal{K}}{\pi r_0^{3/2}}
-(2\ell+1)|\Delta r|\frac{(r_0-M)}{r_0^{5/2}}
\right.\\
&\qquad\qquad\left.+\Delta r\left(\frac{2[r_0^2-3Mr_0+2M^2]\mathcal{E}}{\pi r_0^{7/2}f_0^{1/2}}-\frac{4[r_0^2-3Mr_0+4M^2]\mathcal{K}}{\pi r_0^{7/2}f_0^{1/2}}\right)\right]+O(\Delta r^2).
\end{align}
\end{widetext}

The final task is to express the singular field projections in terms of the original \((t,r,\theta,\phi)\) coordinates. This is accomplished, in part, by reversing the rotation performed in Eqs.~\eqref{rotation} through use of the Wigner-D matrix \(D^\ell_{\,m,m'}\) defined in App.~\ref{rotations},
\beq
\label{ASfullrotation}
\widehat{\tA}_{\text{L},\tS}^{\ell m}=\sum_{m'=-\ell}^\ell D^\ell_{\,m,m'}\left(\pi,\frac\pi2,\frac\pi2\right)\tA_\tS^{\ell m'}.
\eeq

To recover the \(r\)-dependence, we simply substitute the definition of \(\Delta r=r-r_0\). Finally, the singular field projections are evaluated at \(\Delta t=t=0\) in the (\(\Theta,\Phi\)) coordinates. After rotation back to the original Schwarzschild coordinates, the singular field must obey the helical symmetry of the physical spacetime, as it is an approximation of the particular solution to Eq.~\eqref{linearizedEFE}. We then attribute the same time dependence given to the retarded metric perturbation, Eq.~\eqref{trseparable}, written in full as,
\beq
\label{Strseparable}
\tA_{\text{L},\tS}^{\ell m}(t,r)=e^{-i\omega_m t}\sum_{m'=-\ell}^\ell D^\ell_{\,m,m'}\left(\pi,\frac\pi2,\frac\pi2\right)\tA_\tS^{\ell m'}(0,r).
\eeq

The construction of the singular field in this paper is identical to \cite{WardellPRD92} with two additional considerations:
\begin{itemize}
\item[(1)] Constructing the gauge-invariants \eqref{alpha}-\eqref{epsilon} requires taking additional radial derivatives of the singular field projections, so terms proportional to \(\Delta r^2\) are necessary, which were suppressed in the analysis above (for brevity) and in \cite{WardellPRD92}.
\item[(2)] The even-parity gauge-invariants \eqref{chi}-\eqref{epsilon} involve factors of \(\ell\) (contained in \(\lambda\)), which indicate the presence of additional angular derivatives before the mode decomposition. Therefore, a higher-order expansion in \(m'\) is required for certain modes, with the specific value of \(m'\) for each A--K listed in Table~\ref{mpvalues}.
\end{itemize}
The expressions for the higher-order singular field projections are unwieldy, and as such, they are made available electronically \cite{BarryWebsite}, constructed in the BLS basis. One may recover the higher-order projections of the A--K terms used in this work via App.~\ref{BLSbasis}.


\subsection{Singular Field for \(\ell\ge2\) \label{hSlg2}}
To find the RW/EZ gauge singular field, the gauge-invariant quantities in Eqs.~\eqref{alpha}-\eqref{epsilon} are constructed from the A--K projections of \(h_{ab}^{\text{L},\tS}\). Taking into consideration the time dependence in Eq.~\eqref{Strseparable}, the radial functions of the gauge-invariants are,
\begin{align}
\label{alphaS}\widehat{\alpha}_\tS&=\st{J}-\frac r2\rd{\st{G}},\\
\label{betaS}\widehat{\beta}_\tS&=-\st{C}+\frac{i\omega_m r}2\st{G},\\
\label{chiS}\widehat{\chi}_\tS&=\st{H}-\frac{1}{2f}\st{E}-\frac{\lambda+2}{4f}\st{F}-\frac{r}{2}\rd{\st{F}},\\
\nonumber\widehat{\psi}_\tS&=\frac12\st{K}-\frac{r-3M}{2rf^2}\st{E}-\frac{r}{2f}\rd{\st{E}}\\
\label{psiS}&\;\;\;\;-\frac{(\lambda+2)(r-3M)}{4rf^2}\st{F}-\frac{r(\lambda+2)}{4f}\rd{\st{F}},\\
\nonumber\widehat{\delta}_\tS&=\st{D}-\frac{i\omega_m r}{2f}\st{E}-\frac{r-4M}{rf}\st{B}-r\rd{\st{B}}\\
\label{deltaS}&\;\;\;\;-\frac{i\omega_m[r(\lambda+2)-4(r-3M)]}{4f}\st{F}+\frac{i\omega_m r^2}{2}\rd{\st{F}},\\
\nonumber\widehat{\epsilon}_\tS&=-\frac12\st{A}-\frac{M}{2r}\st{E}+i\omega_m r\st{B}\\
\label{epsilonS}&\;\;\;\;+\frac12\left(\omega_m^2r^2-\frac{M(\lambda+2)}{2r}\right)\st{F}.
\end{align}
The \(\ell m\)-modes of the singular metric perturbation in the EZ gauge are found via Eqs.~\eqref{Aez}-\eqref{Kez}, and in the RW gauge via Eqs.~\eqref{Arw}-\eqref{Krw}. As the above quantities are gauge-invariant and the metric reconstruction requires no integration, the (\(\ell\ge2\)) modes of the EZ and RW gauge singular fields are uniquely fixed.


\subsection{Singular Field for \(\ell=0,1\)}

\subsubsection{\(\ell=0\)}
The Zerilli gauge monopole is gauge-invariant under gauge transformations which respect the HKV symmetry. Performing the gauge transformation outlined in Sec.~\ref{l0ret} on the singular field yields,
\begin{align}
\label{AS0}\widehat{\tA}_{\text{Z},\tS}^{00}&=\widehat{\tA}_{\text{L},\tS}^{00}+\frac{M}{r}\widehat{\tE}_{\text{L},\tS}^{00},\\
\label{KS0}\widehat{\tK}_{\text{Z},\tS}^{00}&=\widehat{\tK}_{\text{L},\tS}^{00}-\frac{(r-3M)}{rf^2}\widehat{\tE}_{\text{L},\tS}^{00}-\frac{r}{f}\frac{\mathrm{d}\widehat{\tE}_{\text{L},\tS}^{00}}{\mathrm{d}r}.
\end{align}
Note that the additional gauge transformation between the Lorenz and Zerilli gauges to ensure asymptotic flatness, Eq.~\eqref{NAFgaugevec}, is naturally included in the regular piece of the gauge vector Eq.~\eqref{SRgaugevec}, for it is proportional to a homogeneous solution the EFEs.

\subsubsection{\(\ell=1\) Odd-Parity}
As discussed in Sec.~\ref{l1oddret}, no gauge transformation is necessary for the odd-parity dipole and the singular field structure remains identical. As such, the singular field for the odd-parity Zerilli dipole is equal to the Lorenz gauge odd-parity dipole,
\beq
\widehat{\tC}^{10}_{\tZ,\tS}=\widehat{\tC}^{10}_{\text{L},\tS}.
\eeq

\subsubsection{\(\ell=1\) Even-Parity}
The even-parity dipole singular field is constructed following the gauge transformation outlined in Sec.~\ref{l01retsol}. The unknown constant \(\widehat{\zeta}^{1m}\) in the even-parity dipole gauge vector is pure gauge and induces a change to the retarded field proportional to a homogeneous solution; we therefore attribute it to the regular piece of the gauge vector in Eq.~\eqref{SRgaugevec}. Additionally, the choice of lower bound in the integral for \(\widehat{\tS}_\text{part}^{1m}\), Eq.~\eqref{Sdefined}, fixes the \(\ell=1\) Zerilli gauge solution recovered after regularization; the choice has been made so that the regularization itself requires no knowledge of the retarded Lorenz gauge solution and no integrals of the singular or retarded field are necessary when evaluated at the particle.

With these choices in place, the singular field contribution to the gauge vector is given by,
\begin{align}
\nonumber\widehat{\tP}^{1m}_\tS&=r\left(\widehat{\tB}^{1m}_\text{L,S}-i\omega_m\widehat{\tS}^{1m}_\text{part,S}\right),\\
\label{l1gaugevectorfinal}\widehat{\tR}^{1m}_\tS&=\frac{r}{2f}\left(\widehat{\tE}^{1m}_{\text{L},\tS}+\frac2r\widehat{\tS}^{1m}_\text{part,S}\right),\\
\nonumber\widehat{\tS}^{1m}_\tS&=\widehat{\tS}^{1m}_\text{part,S},
\end{align}
with,
\beq
\label{Ssingdefined}
\widehat{\tS}^{1m}_\text{part,S}(r)= f^{-1}\int_{r_0}^r\left(f(r')\widehat{\tH}^{1m}_{\text{L,S}}(r')-\frac{1}{2}\widehat{\tE}^{1m}_{\text{L,S}}(r')\right)\,\mathrm{d}r',
\eeq
and the singular field for the Zerilli even-parity dipole is,
\begin{align}
\nonumber\widehat{\tA}^{1m}_{\text{Z},\tS}&=\widehat{\tA}^{1m}_{\text{L},\tS}-2i\omega_mr\widehat{\tB}^{1m}_{\text{L},\tS}+\frac{M}{r}\widehat{\tE}^{1m}_{\text{L},\tS}\\
\label{AS1}&\;\;\;\;+2r\left(\frac{M}{r^3}-\omega_m^2\right)\widehat{\tS}^{1m}_\text{part,S},\\
\nonumber\widehat{\tD}^{1m}_{\text{Z},\tS}&=\widehat{\tD}^{1m}_{\text{L},\tS}-\left(\frac{r-4M}{rf}+r\frac{\mathrm{d}}{\mathrm{d}r}\right)\widehat{\tB}^{1m}_{\text{L},\tS}-\frac{i\omega_mr}{2f}\widehat{\tE}^{1m}_{\text{L},\tS}\\
\label{DS1}&\;\;\;\;+i\omega_m\left(r\frac{\mathrm{d}}{\mathrm{d}r}-\frac{4M}{rf}\right)\widehat{\tS}^{1m}_\text{part,S},\\
\nonumber\widehat{\tK}^{1m}_{\text{Z},\tS}&=\widehat{\tK}^{1m}_{\text{L},\tS}-\left(\frac{r-3M}{rf^2}+\frac{r}{f}\frac{\mathrm{d}}{\mathrm{d}r}\right)\widehat{\tE}^{1m}_{\text{L},\tS}\\
\label{KS1}&\;\;\;\;-\left(\frac{2}{f}\frac{\mathrm{d}}{\mathrm{d}r}-\frac{2M}{r^2f^2}\right)\widehat{\tS}^{1m}_\text{part,S}.
\end{align}

\section{Tensor-Harmonic Regularization \label{THRP}}

The regularization procedure detailed in Sec.~\ref{regularization} requires, as input, the retarded and singular \(\ell\)-modes of the quantity of interest, in this case either the self-force or the redshift invariant. We now construct the tensor-harmonic \(\ell m\)-modes of the redshift invariant and the force from the A--K variables of the metric perturbation in both the EZ and RW gauges. The sum over \(m\) is then done analytically for the singular contributions to construct the tensor-harmonic regularization parameters.


\subsection{Mode Decomposition of \(\bar{u}^t\) and \(\mathcal{F}^r\)\label{forceutdecomp}}

\subsubsection{The Redshift Invariant \(\bar{u}^t\)}
The Detweiler redshift invariant is written for circular orbits in Schwarzschild spacetime as \cite{DetweilerPRD77},
\beq
\bar{u}^t=(1-3M/r_0)^{-1/2}\frac12u^au^bh_{ab}^\text{R}.
\eeq
To perform the regularization outlined in Eq.~\eqref{GItransform}, we require the retarded and singular modes of \(\bar{u}^t\); we find these by extending the definition of the redshift invariant off of the particle's worldline,
\beq
\label{utdef}
\bar{u}^t[h]=(1-3M/r_0)^{-1/2}\frac12\tilde{u}^a\tilde{u}^bh_{ab},
\eeq
for any smooth extension \(\tilde{u}^a\), taken in this work to be the rigid extension used by Barack and Ori~\cite{BarackPRD61}, where the components of the four-velocity are held fixed to their values on the worldline while allowing the metric and Christoffel symbols to vary. It is common to introduce a second gauge-invariant quantity proportional to \(\bar{u}^t\) \cite{SagoPRD78},
\beq
\label{Udef}
\Delta U(x)\equiv\tilde{u}^a\tilde{u}^bh_{ab},
\eeq
and to perform the regularization on \(\Delta U\), recovering \(\bar{u}^t\) afterwards via,
\beq
\label{uTreg}
\bar{u}^t_\tR=(1-3M/r_0)^{-1/2}\frac12\lim_{x\rightarrow x_0}\left[\Delta U^\text{ret}-\Delta U^\tS\right](x).
\eeq
We now find the mode decomposition of \(\Delta U\) in each gauge, as constructed from the tensor-harmonic modes of \(h_{ab}\) and evaluated at the particle.

In the EZ gauge, the even- and odd-parity components are constructed from Eq.~\eqref{eq:hina-k} and Eqs.~\eqref{Aez}-\eqref{Kez} for \(\ell\ge 2\),
\pagebreak
\begin{align}
\nonumber\Delta U^{\text{EZ},\ell m}_\text{even}(x_0)&=\frac{\mathcal{E}^2}{f_0^2}\widehat{\tA}^{\ell m}_\text{EZ}\;Y_{\ell m}\left(\frac{\pi}{2},0\right)\\
\label{Uezeven}&=-\frac{2 r_0}{r_0-3M}\widehat{\epsilon}^{\ell m}\;Y_{\ell m}\left(\frac\pi2,0\right),\\
\nonumber\\
\nonumber\Delta U^{\text{EZ},\ell m}_\text{odd}(x_0)&=-\frac{2\mathcal{E}\mathcal{L}}{r_0f_0}\widehat{\tC}^{\ell m}_\text{EZ}\;\partial_\theta Y_{\ell m}\left(\frac{\pi}{2},0\right)\\
\label{Uezodd}&=\frac{2r_0^2\Omega}{r_0-3M}\widehat{\beta}^{\ell m}\partial_\theta Y_{\ell m}\left(\frac\pi2,0\right),
\end{align}
written in terms of the gauge-invariants introduced in Sec.~\ref{GIs} and substituting in the definitions of the specific energy and angular momentum from Eqs.~\eqref{energymomentum}. The \(\ell m\)-modes of \(\Delta U\) in the RW gauge are similarly constructed via Eqs.~\eqref{Arw}-\eqref{Krw},
\begin{align}
\nonumber\Delta U^{\text{RW},\ell m}_\text{even}(x_0)&=\left[\frac{\mathcal{E}^2}{f_0^2}\widehat{\tA}^{\ell m}_\text{RW}+\frac{\mathcal{L}^2}{r_0^2}\widehat{\tE}^{\ell m}_\text{RW}\right]Y_{\ell m}\left(\frac{\pi}{2},0\right)\\
\label{Urweven}&=-\frac{2 r_0}{r_0-3M}\widehat{\epsilon}^{\ell m}\;Y_{\ell m}\left(\frac\pi2,0\right),\\
\nonumber\\
\nonumber\Delta U^{\text{RW},\ell m}_\text{odd}(x_0)&=-\frac{2\mathcal{E}\mathcal{L}}{r_0f_0}\widehat{\tC}^{\ell m}_\text{RW}\;\partial_\theta Y_{\ell m}\left(\frac{\pi}{2},0\right)\\
\label{Urwodd}&=\frac{2r_0^2\Omega}{r_0-3M}\widehat{\beta}^{\ell m}\partial_\theta Y_{\ell m}\left(\frac\pi2,0\right).
\end{align}
The gauge-invariance of \(\Delta U^{\ell m}\) at the particle for the \(\ell\ge2\) modes is now manifestly apparent by comparing the even- and odd-parity contributions constructed in each gauge. One may perform a similar exercise starting with the metric components in the Lorenz gauge, and the expressions for \(\Delta U^{\tL,\ell m}\) reduce to Eqs.~\eqref{Uezeven} and \eqref{Uezodd} for even- and odd-parity, respectively.

To construct the tensor-harmonic modes of \(\Delta U^{\ell m}\) for \(\ell<2\), we turn to the explicit expressions for the A--K variables in the Zerilli gauge outlined in Sec.~\ref{l01retsol}. For the monopole \(\ell=0\), the only non-vanishing contribution to \(\Delta U\) arises from \(\widehat{\tA}^{00}_\tZ\),
\begin{align}
\label{Ul0}\Delta U^{\tZ,00}(x_0)&=\frac{\mathcal{E}^2}{f_0^2}\widehat{\tA}^{00}_\tZ\,Y_{00}\left(\frac\pi2,0\right).
\end{align}
Here we see that \(\Delta U^{\tZ,00}\) inherits its gauge-invariance from \(\widehat{\tA}^{00}_\tZ\), which is gauge-invariant under helically-symmetric gauge transformations as discussed immediately following Eqs.~\eqref{A0} and \eqref{K0}. One may indeed consider \(\widehat{\tA}^{00}_\tZ\) to be proportional to the \(\ell=0\) reduction of \(\widehat{\epsilon}^{\ell m}\) as defined in Eq.~\eqref{epsilon},
\beq
\widehat{\tA}^{00}_\tZ=-2\widehat{\epsilon}^{00},
\eeq
 given that the vector and tensor contributions to \(\widehat{\epsilon}^{\ell m}\) vanish identically for \(\ell=0\), in which case Eq.~\eqref{Ul0} is equivalent to Eq.~\eqref{Uezeven}.

The dipole \(\ell=1\) contributions are found to be,
\begin{align}
\label{Ul1even}\Delta U^{\text{Z},1 m}_\text{even}(x_0)&=\frac{\mathcal{E}^2}{f_0^2}\widehat{\tA}^{1 m}_\text{Z}Y_{1 m}\left(\frac{\pi}{2},0\right),\\
\label{Ul1odd}\Delta U^{\text{Z},1 0}_\text{odd}(x_0)&=-\frac{2\mathcal{E}\mathcal{L}}{r_0f_0}\widehat{\tC}^{1 0}_\text{Z}\partial_\theta Y_{1 0}\left(\frac{\pi}{2},0\right).
\end{align}
Again, \(\widehat{\tA}^{1m}_\tZ\) is invariant under helically-symmetric gauge transformations at the particle via Eq.~\eqref{deltaAZ}, and may be thought of as the \(\ell=1\) reduction of \(\widehat{\epsilon}^{\ell m}\),
\beq
\widehat{\tA}^{1m}_\tZ=-2\widehat{\epsilon}^{1m},
\eeq
and thus Eqs.~\eqref{Ul1even} and \eqref{Uezeven} are equivalent. Note that this correspondence between \(\widehat{\tA}^{1m}_\tZ\) and \(\widehat{\epsilon}^{1m}\) does not hold off the worldline and for radial derivatives of these functions; radial derivatives of \(\widehat{\epsilon}^{\ell m}\) remain gauge-invariant, but radial derivatives of \(\widehat{\tA}^{1m}_\tZ\) depend on choice of gauge, even at the particle, as seen in Eq.~\eqref{zerillihomogeneous}.

\subsubsection{The Force \(\mathcal{F}^r\)}
We next turn to the mode decomposition of the self-force. The full expression for the gravitational self-force is given by Eq.~\eqref{forcehR}. We are interested specifically in regularizing the radial component of the force, which reduces to a simple form for circular orbits in terms of the retarded metric perturbation,
\beq
\mathcal{F}^r[h]=\frac f2\tilde{u}^a\tilde{u}^b\partial_rh_{ab},
\eeq
using the same four-velocity extension as in Eq.~\eqref{utdef}. The even- and odd-parity contributions to the force in the EZ gauge are found for \(\ell\ge2\),
\begin{align}
\nonumber\mathcal{F}^{r,\ell m}_\text{EZ,even}(x_0)&=\frac{\mathcal{E}^2}{2f_0}\partial_r\widehat{\tA}^{\ell m}_\text{EZ}\;Y_{\ell m}\left(\frac{\pi}{2},0\right)\\
\label{forceEZeven}&=-\frac{r_0f_0}{r_0-3M}\partial_r\widehat{\epsilon}^{\ell m}\;Y_{\ell m}\left(\frac\pi2,0\right),\\
\nonumber\\
\nonumber\mathcal{F}^{r,\ell m}_\text{EZ,odd}(x_0)&=-\frac{\mathcal{E}\mathcal{L}}{r_0^2}\left[\widehat{\tC}^{\ell m}_\text{EZ}+r_0\partial_r\widehat{\tC}^{\ell m}_\text{EZ}\right]\partial_\theta Y_{\ell m}\left(\frac{\pi}{2},0\right)\\
\label{forceEZodd}&=\frac{r_0f_0\Omega}{r_0-3M}\left[\widehat{\beta}^{\ell m}+r_0\partial_r\widehat{\beta}^{\ell m}\right]\partial_\theta Y_{\ell m}\left(\frac\pi2,0\right),
\end{align}
expressed in terms of the gauge invariants \(\widehat{\epsilon}^{\ell m}\) and \(\widehat{\beta}^{\ell m}\). In the RW gauge,
\begin{align}
\nonumber\mathcal{F}^{r,\ell m}_\text{RW,even}(x_0)&=\left[\frac{\mathcal{E}^2}{2f_0}\partial_r\widehat{\tA}^{\ell m}_\text{RW}\right.\\
\nonumber&\;\;\;\left.+\frac{f_0\mathcal{L}^2}{2r_0^3}\left(r_0\partial_r\widehat{\tE}^{\ell m}_\text{RW}+2\widehat{\tE}^{\ell m}_\text{RW}\right)\right]Y_{\ell m}\left(\frac{\pi}{2},0\right)\\
\nonumber&=-\frac{f_0}{r_0(r_0-3M)}\\
\label{forceRWeven}&\quad\times\left[r_0^2\partial_r\widehat{\epsilon}^{\ell m}+3Mf_0\widehat{\chi}^{\ell m}\right]Y_{\ell m}\left(\frac\pi2,0\right),\\
\nonumber\\
\nonumber\mathcal{F}^{r,\ell m}_\text{RW,odd}(x_0)&=-\frac{\mathcal{E}\mathcal{L}}{r_0^2}\left[\widehat{\tC}^{\ell m}_\text{RW}+r_0\partial_r\widehat{\tC}^{\ell m}_\text{RW}\right]\partial_\theta Y_{\ell m}\left(\frac{\pi}{2},0\right)\\
\label{forceRWodd}&=\frac{r_0f_0\Omega}{r_0-3M}\left[\widehat{\beta}^{\ell m}+r_0\partial_r\widehat{\beta}^{\ell m}\right]\partial_\theta Y_{\ell m}\left(\frac\pi2,0\right).
\end{align}
The even-parity contributions to the force in the EZ and RW gauges, Eqs.~\eqref{forceEZeven} and \eqref{forceRWeven} respectively, differ by a term proportional to \(\widehat{\chi}^{\ell m}\), while the odd-parity contributions to the force both reduce to identical expressions involving the gauge-invariant \(\widehat{\beta}^{\ell m}\). Further, when constructed in the Lorenz gauge, the odd-parity component of the force exactly matches Eq.~\eqref{forceRWodd}, indicating that the odd-parity contributions to the force are gauge-invariant for circular orbits under the gauge transformations taking the Lorenz gauge to the EZ or RW gauges. This invariance of the odd-parity component of the force is investigated further below.

The low modes of the force are calculated in the Zerilli gauge. For \(\ell=0\) the force is,
\beq
\label{forcel0}\fF^{r,00}_\tZ(x_0)=\frac{\mathcal{E}^2}{2f_0}\partial_r\widehat{\tA}^{00}_\text{Z}Y_{00}\left(\frac{\pi}{2},0\right),
\eeq
which, similarly to \(\Delta U^{\tZ,00}\), is equivalent to Eq.~\eqref{forceEZeven}, and the \(\ell=1\) contributions to the force are given by,
\begin{align}
\label{forcel1even}\mathcal{F}^{r,1 m}_\text{Z,even}(x_0)&=\frac{\mathcal{E}^2}{2f_0}\partial_r\widehat{\tA}^{1 m}_\text{Z}Y_{1 m}\left(\frac{\pi}{2},0\right),\\
\label{forcel1odd}\mathcal{F}^{r,10}_\text{Z,odd}(x_0)&=-\frac{\mathcal{E}\mathcal{L}}{r_0^2}\left[\widehat{\tC}^{10}_\text{Z}+r_0\partial_r\widehat{\tC}^{10}_\text{Z}\right]\partial_\theta Y_{10}\left(\frac{\pi}{2},0\right).
\end{align}

\subsection{Tensor-Harmonic Regularization Parameters}

We now construct the singular contributions to the redshift invariant and the force, and perform the \(m\)-sum analytically to recover the tensor-harmonic regularization parameters introduced in Eq.~\eqref{modesumregT}.

Beginning with the gauge-invariant \(\Delta U\), the singular contributions to the \(\ell\)-modes are determined by,
\beq
\label{Umsum}
\Delta U^\ell_\text{RW,S}=\sum_{m=-\ell}^\ell\Delta  U^{\ell m}_\text{RW,S}(x_0),
\eeq
with
\beq
\label{US}
\Delta U^{\ell m}_\text{RW,S}(x_0)=\lim_{r\rightarrow r_0}\left.\left\{\Delta U^{\text{RW},\ell m}_\text{even}(x)+\Delta U^{\text{RW},\ell m}_\text{odd}(x)\right\}\right\rvert_{\subalign{\theta&=\pi/2\\\phi&=0}}
\eeq
and the terms \(\Delta U^{\text{RW},\ell m}_\text{even}(x)\) and \(\Delta U^{\text{RW},\ell m}_\text{odd}(x)\) are constructed via Eqs.~\eqref{Uezeven} and \eqref{Uezodd} for even- and odd-parity, respectively, from the singular gauge invariants constructed in Eqs.~\eqref{alphaS}-\eqref{epsilonS}. The gauge invariants necessary for the construction of \(\Delta U\) do not involve radial derivatives of the singular field (see Eqs.~\eqref{betaS} and \eqref{epsilonS}). Thus, the singular modes of \(\Delta U\) are continuous across the orbit in all gauges and the limit in Eq.~\eqref{US} does not have directional dependence.

The \(m\)-sum is performed in the original Schwarzschild coordinates, unlike in \cite{WardellPRD92} where the sum is performed over \(m'\) in the rotated coordinates (\(\Theta,\Phi\)). Explicit factors of \(m\) have been introduced into the singular field via the time derivatives in Eqs.~\eqref{alphaS}-\eqref{epsilonS}, and so we perform the sum over azimuthal modes in the unrotated frame. Performing the sum over \(m\) analytically was addressed in \cite{NakanoPRD68}, and we describe its solution in App.~\ref{tensorbasis}. A method to perform the \(m\)-sum in the rotated frame has also been outlined by Miller \textit{et al.}~\cite{MillerPRD94}.

After taking the \(m\)-sum in Eq.~\eqref{Umsum}, we recover \(\Delta U^\ell_\text{RW,S}\) as an expansion in \(\ell\), which we write as two terms,
\begin{align}
\label{USL}
\Delta U^\ell_\text{RW,S}&=\Delta U^\ell_{[0]}+\Delta U^\ell_{[2]},
\end{align}
following the notation for the scalar-harmonic regularization parameters introduced in \cite{HeffernanPRD86}, where a term \(\Delta U^\ell_{[n]}\) scales as \(O(\ell^{-n})\). We find,

\begin{align}
\label{U0} \Delta U^\ell_{[0]}&=\frac{4\mu}{\pi(r_0^2+\mathcal{L}^2)^{1/2}}\hat{\mathcal{K}},\\
\nonumber \Delta U^\ell_{[2]}&=\frac{1}{(2\ell-1)(2\ell+3)}\frac{6\mu}{\pi r_0^2(r_0-3M)^{1/2}(r_0-2M)^{1/2}}\\
\nonumber&\quad\times\left[(5r_0^2-31Mr_0+32M^2)\hat{\mathcal{K}}\right.\\
\label{U2} &\quad\qquad\left.-(r_0-2M)(5r_0-11M)\hat{\mathcal{E}}\right].
\end{align}

Our result for \(\Delta U^\ell_{[0]}\) is identical to the leading-order tensor-harmonic regularization parameter for \(\Delta U\) derived in \cite{WardellPRD92}, the term proportional to \(\ell^{-1}\) vanishes identically, and the result for \(\Delta U^\ell_{[2]}\) is new for tensor-harmonic modes. For the purposes of this work, \(\Delta U^\ell_{[2]}\) acts to accelerate the convergence of the regularization in a similar way to the accelerated convergence techniques used in scalar-harmonic self-force regularization \cite{DetweilerPRD67,HeffernanPRD86}, as visualized in Fig.~\ref{uregfalloff} and detailed further in Sec.~\ref{results}.

To construct the regularization parameters for the force, we perform the \(m\)-sum as outlined in Eq.~\eqref{forcelmodes},
\beq
\fF^{r,\ell\pm}_\tS=\left.\sum_{m=-\ell}^\ell \lim_{r\rightarrow r_0^\pm}\fF^{r,\ell m}[h^\text{RW,S}]\right.\rvert_{\subalign{\theta&=\pi/2\\\phi&=0}},
\eeq
where the \(\ell m\)-modes of the force are calculated from the singular gauge invariants, Eq.~\eqref{alphaS}-\eqref{epsilonS}, using Eqs.~\eqref{forceEZeven} and \eqref{forceEZodd} in the EZ gauge and Eqs.~\eqref{forceRWeven} and \eqref{forceRWodd} in the RW gauge. The modes for \(\ell<2\) are found in the Zerilli gauge as outlined in Eqs.~\eqref{forcel0}-\eqref{forcel1odd}. After the \(m\)-sum is performed, the \(\ell\)-modes of the singular force separate into two terms, as in Eq.~\eqref{modesumregT},
\beq
\label{singularforcetoregparam}
\fF^{r,\ell\pm}_\tS=(2\ell+1)\fF^{r,\pm}_{[-1]}+\fF^{r}_{[0],\text{RW}},
\eeq
with the leading-order singular contribution given in both the EZ and RW gauges as,
\begin{align}
\nonumber\fF^{r,\pm}_{[-1]}&=\mp\frac{\mu^2}{2r_0^2}\left(1-\frac{3M}{r_0}\right)^{1/2}\pm\left[\frac{2\mu^2M(2M-r_0)}{r_0^{5/2}(r_0-3M)^{3/2}}\right]_{\ell<1}\\
\label{Fn1}&\quad\pm\left[\frac{\mu^2M^2}{2r_0^{5/2}(r_0-3M)^{3/2}}\right]_{\ell<2}.
\end{align}
This term is independent of the choice of EZ or RW gauge, and is identical to the leading-order tensor-harmonic regularization term used for the Lorenz gauge force \cite{WardellPRD92}. We note that this behavior is also observed when regularizing the self-force in the radiation gauge, where the leading-order scalar-harmonic regularization parameters are found to be identical in both the radiation and Lorenz gauges \cite{ShahPRD83}.

The regularization term \(\fF^{r}_{[0],\text{RW}}\) does depend on the choice of gauge. We opt to write the sub-leading regularization parameters following Nakano \etal \cite{NakanoPRD68}, where the following regularization parameters are defined for all \(\ell\) and adjustments due to the \(\ell<2\) modes are written as separate corrections. The sub-leading tensor-harmonic regularization parameters for the EZ and RW gauges are finally given by,

\begin{align}
\nonumber\mu^{-2}\fF^{r}_{[0],\text{EZ}}&=\frac{(r_0-2M)^{1/2}}{\pi  r_0^3 (r_0-3 M)^{3/2}} \left[ \left(33 M^2-18 M r_0+r_0^2\right)\hat{\mathcal{E}}\right.\\
\label{F0EZ}&\left.\qquad-2 \left(18 M^2-9M
   r_0+r_0^2\right)\hat{ \mathcal{K}}\right],\\
   \nonumber\\
\label{F0RW}\mu^{-2} \fF^{r}_{[0],\text{RW}}&=\frac{(r_0-2M)^{1/2}(r_0-3M)^{1/2}}{\pi r_0^3}[\hat{\mathcal{E}}-2\hat{\mathcal{K}}].
\end{align}

Looking first to Eq.~\eqref{F0RW}, we note that this term is identical to the Lorenz gauge \(B^r\) parameter for scalar-harmonic regularization and the non-vanishing contribution to \(\fF^r_{[0],\tL}\) in \cite{WardellPRD92}. The only deviation away from the Lorenz-gauge regularization lies in the adjustments made at \(\ell<2\); as these adjustments arise from the difference between the asymptotic, high-\(\ell\) behavior of the singular modes of the force and the local expansion of the singular force, they are naturally attributed to the \(D^r\) parameter in Eq.~\eqref{modesumregT}~\cite{BarackPRD67}, which is found to vanish in the Lorenz gauge but in the RW gauge now takes a non-zero value given in Eq.~\eqref{DRW}.

Thus, regularization may be performed in the RW gauge by using the Lorenz gauge tensor-harmonic regularization parameters with the addition of a non-vanishing \(D^r_\text{RW}\) parameter. The same may not be said of regularization in the EZ gauge: the \(\ell\)-independent contribution to \(\fF^r_{[0],\text{EZ}}\) is not equal to the Lorenz gauge term. Regularization in this gauge requires an adjustment not only to \(D^r_\text{EZ}\) given in Eq.~\eqref{DEZ}, but also an adjustment to the Lorenz gauge \(B^r\) parameter at each \(\ell\). One sees this result more directly when the singular gauge vector between the Lorenz and EZ/RW gauges is constructed, which we now do.

The method outlined above for constructing the force regularization parameters involves first finding the singular gauge invariants in Eqs.~\eqref{alphaS}-\eqref{epsilonS} and then reconstructing the singular contributions to the force directly in each gauge. An equally valid approach to finding the regularization parameters is to explicitly calculate the gauge vector between the Lorenz gauge and the EZ/RW gauges. The force then transforms as in Eq.~\eqref{forcegaugeT}; for a particle traveling along a circular orbit, the radial component of the force transforms under gauge transformations which obey the HKV symmetry as \cite{KeidlPRD82},
\beq
\label{forcegaugetrans}
\mathcal{F}^r_\text{new}(x_0)=\mathcal{F}^r_\text{old}(x_0)-\frac{3\mu Mf_0}{r_0^2(r_0-3M)}\xi^r.
\eeq
To construct the regularization parameters in the EZ/RW gauges, we require the Lorenz gauge tensor-harmonic regularization parameters (found in \cite{WardellPRD92}) and the mode-decomposition of the singular gauge vector introduced in Eq.~\eqref{singularGT}. The gauge-transformed regularization parameters are then given by,
\beq
\fF^{r,\ell\pm}_\text{RW,S}=\fF^{r,\ell\pm}_\text{L,S}-\frac{3\mu Mf_0}{r_0^2(r_0-3M)}\xi^{r,\ell\pm}_\text{RW,S}.
\eeq
 For the \(\ell\ge2\) modes, the radial component of the gauge vector \(\xi^{a}_{\tS}\) is straightforward to find for both gauges \cite{TCW} from the tensor-harmonic modes of the Lorenz gauge singular field,
\begin{align}
\label{EZgvS}\hat{\xi}^{r,\ell m}_\text{EZ,S}(x_0)&=\frac{r_0}{2}\left[\widehat{\tE}_{\tL,\tS}^{\ell m}+\frac{\lambda+2}{2}\widehat{\tF}^{\ell m}_{\tL,\tS}\right]Y_{\ell m}\left(\frac\pi2,0\right),\\
\label{RWgvS}\hat{\xi}^{r,\ell m}_\text{RW,S}(x_0)&=r_0f_0\left[\widehat{\tH}_{\tL,\tS}^{\ell m}-\frac{r_0}{2}\frac{\rmd \widehat{\tF}_{\tL,\tS}^{\ell m}}{\rmd r}\right]Y_{\ell m}\left(\frac\pi2,0\right).
\end{align}
The \(\ell<2\) modes of the gauge transformation require the gauge vector from the Lorenz gauge to the Zerilli gauge, as outlined in Sec.~\ref{l01retsol}, with the monopole contribution given by Eq.~\eqref{l0gaugevecR},
\beq
\hat{\xi}^{r,00}_{\tZ,\tS}(x_0)=\frac{r_0}{2}\widehat{\tE}_{\tL,\tS}^{00}Y_{00}\left(\frac\pi2,0\right).
\eeq
For \(\ell=1\), only even-parity requires a gauge transformation, and the radial component to the gauge vector is given by Eq.~\eqref{Rl1},
\beq
\xi_{\tZ,\tS}^{r,1m}(x_0)=\frac{r_0}{2}\widehat{\tE}_{\tL,\tS}^{1m}Y_{1m}\left(\frac\pi2,0\right).
\eeq
Recall that a choice was made in Sec.~\ref{singfieldconst} to associate the gauge constant \(\widehat{\zeta}^{1m}\) with the regular contribution to the dipole gauge transformation and that \(\widehat{\tS}^{1m}_\text{part}\) vanishes at the orbit.

We list the full expressions for the gauge vectors in App.~\ref{forceregparam}, but the results of this calculation are not surprising and produce the same regularization parameters presented above. In the RW gauge, \(\xi^{r,\ell}_\text{RW,S}\) contains at leading-order terms which scale as \(O(\ell^{-2})\) and vanish when summed from \(\ell=0\) to infinity, plus contributions specifically at \(\ell<2\) that generate \(D^r_\text{RW}\). The EZ gauge vector \(\xi^{r,\ell}_\text{EZ,S}\) scales as a constant at leading-order in \(\ell\), along with terms which vanish in the \(\ell\)-sum and specific contributions at \(\ell<2\). This constant scaling behavior in the \(\ell\)-sum corresponds to a local \(1/s\) singularity in the gauge vector \cite{ShahPRD83} that matches the local analysis performed in App.~\ref{LorenzToEZ}.

\section{Results \label{results}}
We now list the results of our numerical analysis, beginning with the regularization of the redshift invariant. We then calculate the regularized LL force from both the RW and EZ gauge retarded metric perturbations. Finally, we calculate the gauge vector, \(\xi^\text{RW,C}\), from the regularized Lorenz gauge metric perturbation and compare the Lorenz gauge self-force to the forces computed in the EZ and RW gauges.

To ensure that the comparison occurs at the same event in all gauges, we work with an asymptotically flat monopole as discussed in Sec.~\ref{l01retsol}, and evaluate all quantities at the gauge-invariant radius introduced in \cite{DetweilerPRD77},
\beq
R_\Omega\equiv\left(\frac{M}{\Omega^2}\right)^{\frac13}.
\eeq
For quantities which are entirely first-order in \(\mu/M\), e.g., \(\bar{u}^t\), we find that \(\bar{u}^t(R_\Omega)=\bar{u}^t(r_0)+O(\mu^2/M^2)\) \cite{SagoPRD78}.

The regularized redshift invariant \(\bar{u}^{t}_\tR\) is calculated by performing the sum in Eq.~\eqref{modesumI}, subtracting the tensor-harmonic regularization parameters from retarded \(\ell\)-modes,
\beq
\bar{u}^{t}_\tR=(1-3M/R_\Omega)^{-1/2}\frac12\sum_{\ell=0}^{\ell_\text{max}}\left[\Delta U^\ell_\text{RW,ret}-\Delta U_\text{[0]}^\ell-\Delta U^\ell_{[2]}\right],
\eeq
with the \(\ell<2\) modes of \(\Delta U^\ell_\text{RW,ret}\) constructed in the Zerilli monopole and dipole gauges, respectively, for both the RW and EZ gauges.
To calculate the regularized radial component of the self-force in each gauge, we perform the summation,
\beq
\label{modesumregRW}
\fF^r_\text{self}=\sum_{\ell=0}^{\ell_\text{max}}\left[\fF^{r,\ell\pm}_\text{RW,ret}-(2\ell+1)\fF_{[-1]}^{r,\pm}-\fF_{[0],\text{RW}}^r\right]-D^r_\text{RW},
\eeq
where the retarded modes of the force are calculated in each gauge following Sec.~\ref{forceutdecomp}.

\begin{figure}[t!]
\includegraphics[width=0.5\textwidth]{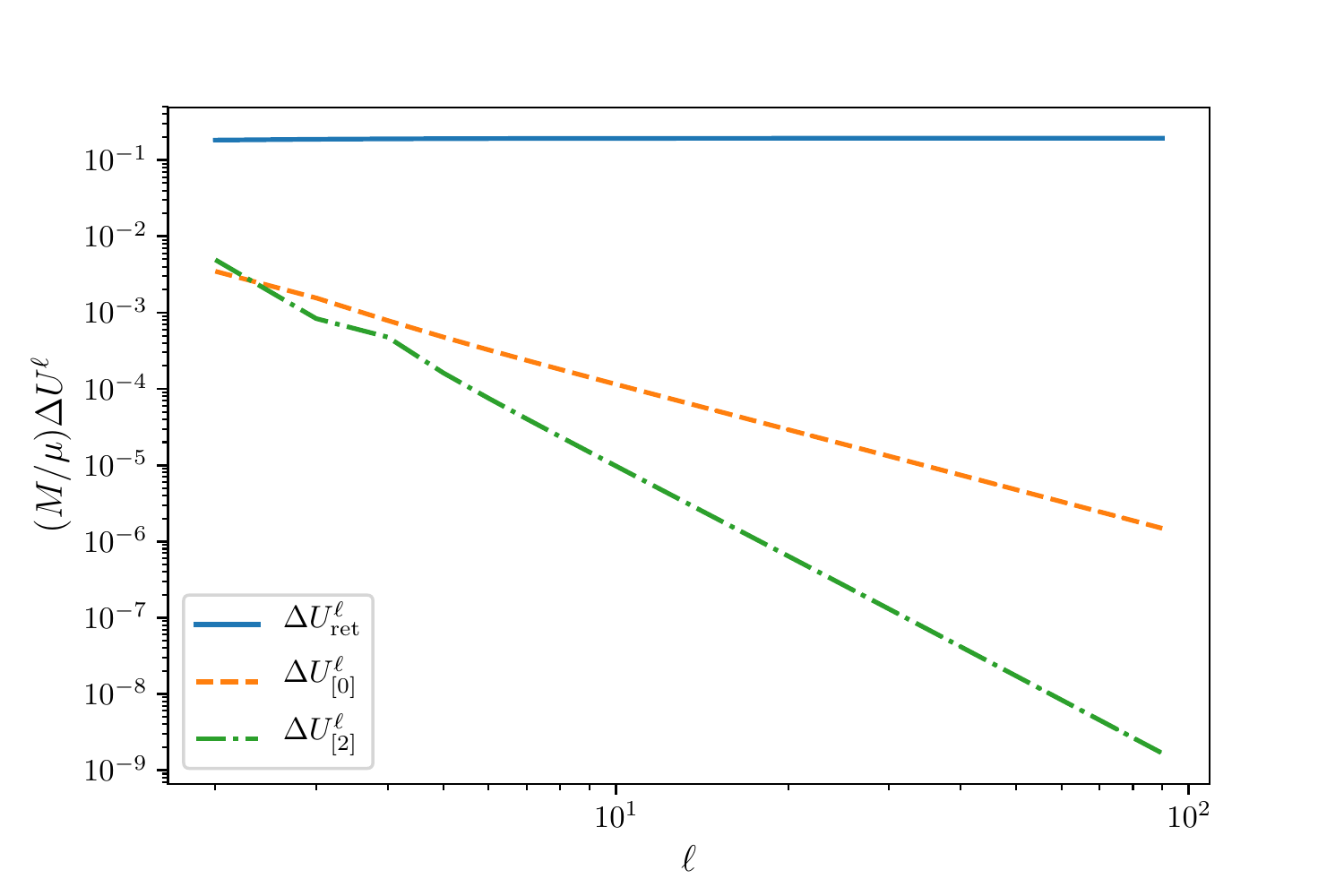}
\caption{We plot the \(\ell\)-modes of \(\Delta U^\ell\) for \(2\le\ell\le90\) on a log-log scale at the orbital radius \(r_0=10M\). \(\Delta U^\ell_\text{ret}\) denotes the unregularized, retarded modes of \(\Delta U^\ell\) constructed in the RW gauge, while \(\Delta U^\ell_{[0]}\) and \(\Delta U^\ell_{[2]}\) correspond to the regularized modes of \(\Delta U^\ell\) after subtracting first \(\Delta U^\ell_{[0]}\) and then \(\Delta U^\ell_{[2]}\) from the retarded modes, respectively, as in Eq.~\eqref{Uregdef}.   \label{uregfalloff}}
\end{figure}

\begin{figure}[t!]
\includegraphics[width=0.5\textwidth]{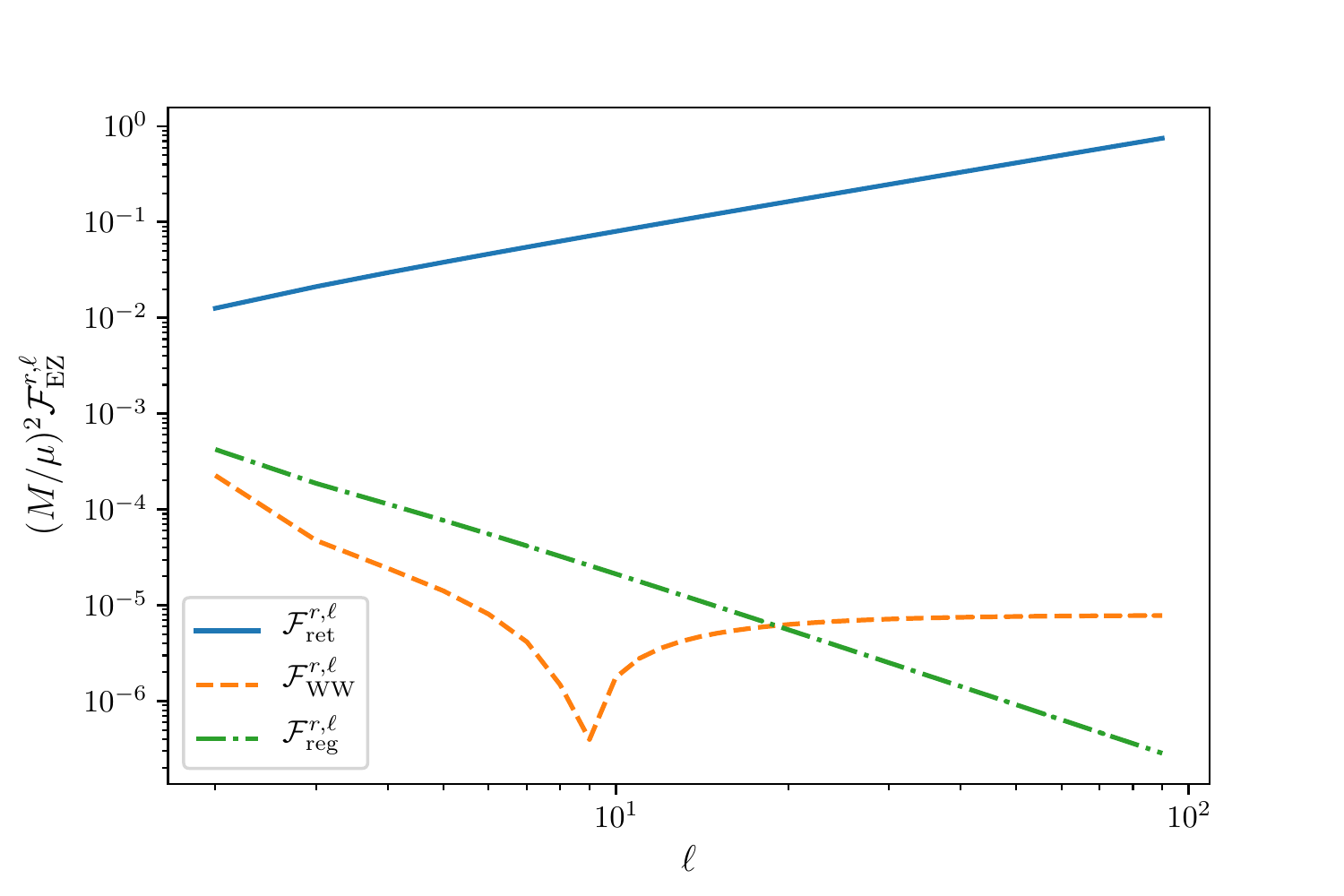}
\caption{The individual \(\ell\)-modes of the EZ gauge self-force are plotted for \(2\le\ell\le90\) on a log-log scale at the orbital radius \(r_0=10M\). The retarded modes of the EZ gauge self-force, \(\fF^{r,\ell}_\text{ret}\), are calculated from inside the orbit and are shown to diverge with \(\ell\). The force \(\fF^{r,\ell}_\text{WW}\) corresponds to the regularization produced when using the low-order, analytic expansions for the singular field published in \cite{WardellPRD92}; this regularization is incomplete and the self-force diverges in the \(\ell\)-sum as \(1/s\). Finally we plot the regularized EZ gauge self-force, \(\fF^{r,\ell}_\text{reg},\) produced in Eq.~\eqref{modesumregRW} using the EZ gauge regularization parameters in Sec.~\ref{THRP}. \label{fregfalloff}}
\end{figure}

To account for the truncation of the sums above at \(\ell_\text{max}\), we introduce a ``tail'' correction \cite{BarackPRD75,ShahPRD86}, for \(\Delta U\) given by,
\beq
\label{Uregtail}
\Delta U_\text{tail}=\sum_{\ell_\text{max}+1}^\infty \Delta U^\ell_\text{res},
\eeq
with \(\Delta U^{\ell}_\text{res}\) defined as,
\beq
\label{Uregdef}\Delta U^\ell_\text{res}\equiv \Delta U^\ell_\text{RW,ret}-\Delta U^\ell_{[0]}-\Delta U^\ell_{[2]}.
\eeq
\(\Delta U^\ell_\text{res}\) is found by numerically fitting the \(\ell\)-falloff of the  \(O(\epsilon^0)\) and higher contributions to the residual, plotted in Fig.~\ref{uregfalloff}, assuming it has the form given by the Ansatz,
\begin{equation}
\label{highorder}
\Delta U^{\ell}_{[4+]}\sim\sum_{k=2}^{k_\text{max}}\frac{\Delta U_{[2k]}}{P_{2k}(\ell)},
\end{equation}
where each \(P_{2k}(\ell)\) is a polynomial of order \(\ell^{2k}\) chosen such that each term in the sum Eq.~\eqref{highorder} vanishes when summed from \(\ell=0\) to infinity (and thus does not formally contribute to the self-force), and \(\{\Delta U_{[2k]}\}_{k=2}^{k_\text{max}}\) are constant parameters. We use the polynomials given by,
\begin{equation}
P_{2k}(\ell)=\prod_{k'=0}^{k}(2\ell-2k'-1)(2\ell+2k'+3),
\end{equation}
which we note are naturally found in the accelerated term Eq.~\eqref{U2} for \(k=0\).

To accelerate convergence of the regularized self-force, we assume a similar form for the residual,
\begin{equation}
\fF^{r,\ell}_\text{res}\equiv\fF^{r,\ell\pm}_\text{RW,ret}-(2\ell+1)\fF_{[-1]}^{r,\pm}-\fF_{[0],\text{RW}}^r,
\end{equation}
and fit the data to the Ansatz,
\begin{equation}
\fF^{r,\ell}_{[2+]}\sim\sum_{k=1}^{k_\text{max}}\frac{\fF^r_{[2k]}}{P_{2k}(\ell)},
\end{equation}
beginning here at \(k=1\) to match the \(\ell\)-falloff of the residual data. The acceleration to the convergence is then seen as \cite{ShahPRD86},
\begin{align}
\nonumber\mathcal{F}^r_\text{R}&=\sum_{\ell=0}^{\ell_\text{max}}\left[\mathcal{F}^{r,\ell}_\text{ret}-(2\ell+1)\mathcal{F}^{r,\ell}_{[-1]}-\mathcal{F}^{r,\ell}_{[0]}\right]\\
\label{accelalt}&\quad+\sum_{\ell_\text{max}+1}^\infty\sum_{k=1}^{k_\text{max}}\frac{\mathcal{F}^r_{[2k]}}{P_{2k}(\ell)}+O\left(\ell^{-(2k_\text{max}+1)}\right).
\end{align}

The final results for \(\bar{u}^{t}_\tR\) are tabulated in Table~\ref{uttable} for a variety of orbital radii, compared against the results of Dolan \etal \cite{DolanPRD91}. The results for the regularized self-force computed in the EZ and RW gauges are given in Table~\ref{forcetable} for a variety of orbital radii.

\begin{figure}
\includegraphics[width=0.5\textwidth]{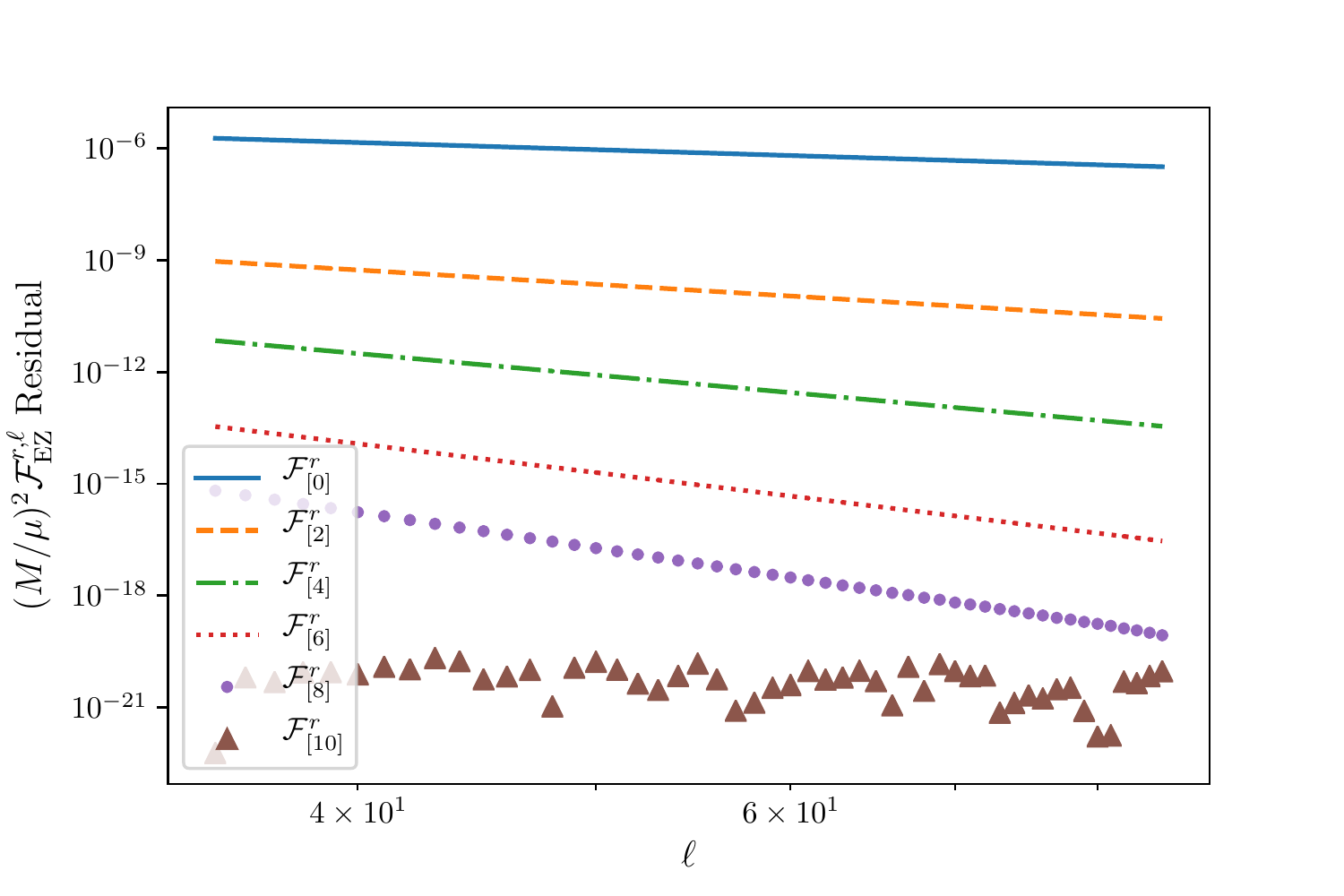}
\caption{Absolute value of the residual after subtraction of each successive regularization term from the EZ gauge self-force versus \(\ell\) on a log-log scale from \(\ell_\text{min}=35\) to \(\ell_\text{max}=85\) for \(r_0=10M\) and \(k_\text{max}=5\). \label{accelfig}}
\end{figure}

\subsection{Comparison to Lorenz Gauge Force}
As a check of our results, we now calculate the gauge transformation between the regularized self-force in the Lorenz gauge and each of the EZ and RW gauges by computing the regular gauge vector in Eq.~\eqref{SRgaugevec}. We choose to begin in the Lorenz gauge and work to find the gauge transformation to the EZ/RW gauges; this choice is a matter of convenience, since the gauge transformation from any gauge to the EZ/RW gauges is relatively simple to construct using tensor-harmonic modes \cite{TCW}, while the gauge transformation from the RW gauge to the Lorenz gauge is not \cite{Berndtson:2009hp}. The task is further simplified by the need for only the radial component of the gauge vector, as seen by the transformation properties of the force for circular orbits in Eq.~\eqref{forcegaugetrans}.

To start, the regularized Lorenz gauge metric perturbation, \(h_{ab}^{\tL,\tR},\) is computed using the \textit{effective source} regularization techniques outlined in \cite{WardellPRD92}. The authors were given this numerical data from Niels Warburton~\cite{WarburtonComm}, decomposed into BLS-basis tensor-harmonic modes. From this numerical data, the regular gauge vector is constructed mode-by-mode following Eq.~\eqref{SRgaugevec}.

 For \(\ell\ge2\), the radial component of the regular gauge vector \(\xi^{a}_{\tC}\) is constructed identically to the singular gauge vector \(\xi^a_\tS\), replacing the singular A--K components of the Lorenz gauge metric perturbation with the regularized components,
\begin{align}
\label{EZgvC}\hat{\xi}^{r,\ell m}_\text{EZ,C}(x_0)&=\frac{r_0}{2}\left[\widehat{\tE}_{\tL,\tR}^{\ell m}+\frac{\lambda+2}{2}\widehat{\tF}^{\ell m}_{\tL,\tR}\right]Y_{\ell m}\left(\frac\pi2,0\right),\\
\label{RWgvC}\hat{\xi}^{r,\ell m}_\text{RW,C}(x_0)&=r_0f_0\left[\widehat{\tH}_{\tL,\tR}^{\ell m}-\frac{r_0}{2}\frac{\rmd \widehat{\tF}_{\tL,\tR}^{\ell m}}{\rmd r}\right]Y_{\ell m}\left(\frac\pi2,0\right).
\end{align}
with monopole and dipole contributions given by,
\begin{align}
\widehat{\xi}^{r,00}_{\tZ,\tC}(x_0)&=\frac{r_0}{2}\widehat{\tE}_{\tL,\tR}^{00}Y_{00}\left(\frac\pi2,0\right),\\
\label{gvl1}\widehat{\xi}_{\tZ,\tC}^{r,1m}(x_0)&=\left[\frac{r_0}{2}\widehat{\tE}_{\tL,\tR}^{1m}+f_0^{-1}\widehat{\zeta}^{1m}\right]Y_{1m}\left(\frac\pi2,0\right),
\end{align}
where \(\widehat{\zeta}^{1m}\) is the gauge constant used to specify the specific retarded Zerilli dipole gauge used for the EZ and RW gauge forces, calculated from the \textit{retarded} Lorenz gauge metric perturbation using either Eq.~\eqref{zetaD} or \eqref{zetaK}. The verification of Eq.~\eqref{l1verify}, along with the comparison between the analytic Zerilli dipole metric perturbation and the gauge-transformed Lorenz gauge dipole perturbation provides a check on the numerical accuracy of the Lorenz gauge numerical data produced in \cite{WardellPRD92}, listed in Table~\ref{Ldatacheck}. The full gauge vector is recovered by summing over modes,

\begin{align}
\nonumber\widehat{\xi}^r_\text{RW,C}(x_0)&=\widehat{\xi}^{r,00}_{\tZ,\tC}+\sum_{m=-1}^1\widehat{\xi}^{r,1m}_{\tZ,\tC}\\
&\quad+\sum_{\ell=2}^{\ell_\text{max}}\sum_{m=-\ell}^\ell \widehat{\xi}^{r,\ell m}_\text{RW,C}+\widehat{\xi}^r_\text{tail,RW},
\end{align}
where we have introduced a tail contribution to compensate for the truncated \(\ell\)-sum, defined as above for \(\fF^r_\text{tail}\) and fit to the \(\ell\)-falloff of \(\widehat{\xi}^{r,\ell}_\text{RW,C}\).

Finally, the retarded Lorenz gauge metric begins with a monopole which is not asymptotically flat (see Sec.~\ref{l01retsol}). Adjusting to an asymptotically flat monopole requires the gauge vector given in Eq.~\eqref{NAFgaugevec} which does not obey the HKV symmetry; its contribution to the change in the force must be calculated separately from Eq.~\eqref{forcegaugetrans} using \(\xi^\text{NAF}_a\) \cite{KeidlPRD82},
\beq
\fF^r_\text{AF}(x_0)=\fF^r_\text{NAF}(x_0)+\frac{2\mu^2 M^2f_0}{r_0^{3/2}(r_0-3M)^{3/2}}.
\eeq
Combining these two gauge contributions, the final result of the gauge transformation is,
\begin{align}
\nonumber\fF^r_{\tL\rightarrow\text{RW,R}}(x_0)&=\fF^r_\text{L,R}(x_0)-\frac{3\mu Mf_0}{r_0^2(r_0-3M)}\xi^r_\text{RW,C}(x_0)\\
&\quad+\frac{2\mu^2 Mf_0}{r_0^{3/2}(r_0-3M)^{3/2}}.
\end{align}
We compare the gauge-transformed Lorenz gauge force against the self-forces computed in the EZ and RW gauges in Table~\ref{lorenzEZRWcompare}.

\begin{table}[htbp]
\caption{\label{Ldatacheck} We check the accuracy of the Lorenz gauge numerical data produced in \cite{WardellPRD92} against the analytic Zerilli gauge dipole metric perturbation by performing the gauge transformation and evaluating the relative difference at \(r_0=6M\) and \(r_0=10M\). We also verify that the two gauge constants in Eq.~\eqref{l1verify} coincide for circular orbits. All values are given as relative differences, \(\Delta\tK=|\tK_{\tL\rightarrow\tZ}/\tK_\tZ-1|,\) except for the A term which vanishes in the Zerilli gauge at the particle.}
\begin{ruledtabular}
\begin{tabular}{l l l}
&\(r_0=6M\) &\(r_0=10M\) \\
\hline\\
\((M/\mu)\tA_{\tL\rightarrow\tZ}\)&\(2\times10^{-28}\)&\(5\times10^{-27}\)\\
\(\Delta\partial_r\tA\)&\(8\times 10^{-28}\)&\(2\times10^{-26}\)\\
\(\Delta\tD\)&\(1\times10^{-27}\)&\(1\times 10^{-25}\)\\
\(\Delta\partial_r\tD\)&\(2\times 10^{-27}\)&\(2\times 10^{-25}\)\\
\(\Delta\tK\)&\(2\times10^{-27}\)&\(2\times10^{-25}\)\\
\(\Delta\partial_r\tK\)&\(2\times10^{-27}\)&\(1\times10^{-25}\)\\
\(\Delta\widehat{\zeta}\)&\(7\times10^{-27}\)&\(6\times10^{-25}\)
\end{tabular}
\end{ruledtabular}
\end{table}


\section{Conclusions}
In this work, we produce results for the regularized gravitational self-force computed in the RW and EZ gauges for a circular orbit in the Schwarzschild spacetime, and compare directly our values of these forces to the Lorenz gauge self-force via an explicit gauge transformation. Our numerical implementation allows for the fast and efficient calculation of the first-order self-force from the Regge-Wheeler and Zerilli master functions in the RW gauge itself for circular orbits, which has heretofore not been done.

The results presented here fill a gap in the literature for self-force regularization in the RW and EZ gauges at first-order. They also act as a step toward the development of a framework for gravitational self-force regularization in the RW/EZ gauges at second-order in the perturbation. Thus far, approaches to the second-order analysis have been rooted in the Lorenz gauge (see e.g. \cite{PoundPRD90}). A general approach to perturbations in the RW gauge at second-order in the mass-ratio was introduced by Brizuela \etal \cite{BrizuelaPRD80}. However, ongoing work to regularize the first-order metric perturbation at spatial infinity \cite{PoundPRD92} and the horizon is necessary before construction of the second-order sources is tractable.

\begin{widetext}

\begin{table}[h]
\caption{\label{lorenzEZRWcompare} We provide a numerical comparison between the self-force computed in the Lorenz gauge from numerical data produced in \cite{WardellPRD92} and the self-force constructed in the RW and EZ gauges from numerical data produced for this work, with \(\ell_\text{max}=60\) for the Lorenz-gauge data. The regularized Lorenz gauge self-force is computed using the methods outlined in \cite{WardellPRD92}, and the gauge vectors \(\xi^r_\text{RW,C}\) and \(\xi^r_\text{EZ,C}\) are calculated from the regularized Lorenz gauge metric perturbation as outlined in Eqs.~\eqref{EZgvC}-\eqref{gvl1}. The computed relative difference is taken to be \(|\fF^r_{\tL\rightarrow\text{RW,R}}/\fF^r_\text{RW,R}-1|\).}
\begin{ruledtabular}
\begin{tabular}{l l l}
&\(r_0=6M\) &\(r_0=10M\) \\
\hline\\
\((M/\mu)^2\fF^r_\text{L,R}\)&\(2.4466497159525\times 10^{-2}\)&\(1.338946946191866\times 10^{-2}\)\\
&&\\
\((M/\mu)\xi^r_\text{RW,C}\)&\(-2.430453614878363\)&\(-2.64575603798078264\)\\
\((M/\mu)^2\fF^r_{\tL\rightarrow\text{RW},\tR}\)&\(8.6934324131015\times 10^{-2}\)&\(2.51925841277599\times 10^{-2}\)\\
\((M/\mu)^2\fF^r_\text{RW,R}\)&\(8.6934324131015\times 10^{-2}\)&\(2.51925841277600\times 10^{-2}\)\\
rel. diff.&\(5.\times10^{-15}\)&\(7.\times10^{-17}\)\\
&&\\
\((M/\mu)\xi^r_\text{EZ,C}\)&\(-2.246307322554369\)&\(-2.56810694018625259\)\\
\((M/\mu)^2\fF^r_{\tL\rightarrow\text{EZ},\tR}\)&\(8.3524207606496\times 10^{-2}\)&\(2.49263586496072\times 10^{-2}\)\\
\((M/\mu)^2\fF^r_\text{EZ,R}\)&\(8.3524207606497\times 10^{-2}\)&\(2.49263586496073\times 10^{-2}\)\\
rel. diff.&\(5.\times10^{-15}\)&\(2.\times10^{-16}\)\\
\end{tabular}
\end{ruledtabular}
\end{table}

\begin{table}[h]
\caption{\label{uttable} Comparison between the regularized \(\bar{u}^t\) from this work using \(\ell_\text{max}=90\) and numerical data presented by Dolan \etal \cite{DolanPRD91} in their Table III, evaluated at the gauge-invariant radius \(R_\Omega\). The uncertainty in this work's data is represented by the first excluded digit and is determined by the error in the numerical data.}
\begin{ruledtabular}
\begin{tabular}{l l l}
\(R_\Omega/M\)&\((M/\mu)\,\)\(\bar{u}^t\) [This Work]&\((M/\mu)\,\)\(\bar{u}^t\) [Dolan \etal]\\
\hline\\
5&\(-4.66652374199560\times 10^{-1}\)&\tabdat{-4.666523741995578}{-1}\\
6&\(-2.960275092900145\times 10^{-1}\)&\tabdat{-2.9602750929001455 }{-1}\\
7&\(-2.208475274322470\times 10^{-1}\)&\tabdat{-2.20847527432247320}{-1}\\
8&\(-1.777197435535924\times 10^{-1}\)&\tabdat{-1.77719743553592433}{-1}\\
9&\(-1.493606089179072\times 10^{-1}\)&\tabdat{-1.49360608917907227}{-1}\\
10&\(-1.291222743920494\times 10^{-1}\)&\tabdat{-1.29122274392049459}{-1}\\
12&\(-1.019355723862671\times 10^{-1}\)&\tabdat{-1.01935572386267132}{-1}\\
14&\(-8.438195340957111\times 10^{-2}\)&\tabdat{-8.43819534095711226}{-2}\\
16&\(-7.205505742934500\times 10^{-2}\)&\tabdat{-7.20550574293450112}{-2}\\
18&\(-6.290189942823900\times 10^{-2}\)&\tabdat{-6.29018994282390090}{-2}\\
20&\(-5.582771860249385\times 10^{-2}\)&\tabdat{-5.58277186024938513}{-2}\\
30&\(-3.577831357182052\times 10^{-2}\)&\tabdat{-3.57783135718205099}{-2}\\
40&\(-2.633967741370485\times 10^{-2}\)&\tabdat{-2.63396774137048419}{-2}\\
50&\(-2.084465653059542\times 10^{-2}\)&\tabdat{-2.08446565305954225}{-2}\\
60&\(-1.724759329267916\times 10^{-2}\)&\tabdat{-1.72475932926791548}{-2}\\
70&\(-1.470964636172172\times 10^{-2}\)&\tabdat{-1.47096463617217204}{-2}\\
80&\(-1.282296057577150\times 10^{-2}\)&\tabdat{-1.28229605757714959}{-2}\\
90&\(-1.136531560741143\times 10^{-2}\)&\tabdat{-1.13653156074114270 }{-2}\\
100&\(-1.020528273002761\times 10^{-2}\)&\tabdat{-1.02052827300276055}{-2}\\
500&\(-2.008040444139764\times 10^{-3}\)&\tabdat{-2.00804044413976405}{-3}\\
1000&\(-1.002005027714143\times 10^{-3}\)&\tabdat{-1.00200502771414297}{-3}\\
5000&\(-2.000800400443024\times 10^{-4}\)&\tabdat{-2.00080040044302370}{-4}\\
\end{tabular}
\end{ruledtabular}
\end{table}

\begin{table}[h]
\caption{\label{forcetable} The gravitational self-force calculated for a variety of radii \(r_0\) in the EZ and RW gauges, with \(\ell_\text{max}=90\). The uncertainty in this work's data is represented by the first excluded digit and is determined by the error in the numerical data.}
\begin{ruledtabular}
\begin{tabular}{l l l}
\(r_0/M\)&\((M/\mu)^2\,\fF^r_\text{EZ}\) &\((M/\mu)^2\,\fF^r_\text{RW}\) \\
\hline
5&\(1.3491385787783\times 10^{-1}\)&\(1.4478678123551\times 10^{-1}\)\\
6&\(8.3524207606497\times 10^{-2}\)&\(8.6934324131015\times 10^{-2}\)\\
7&\(5.7062560017807\times 10^{-2}\)&\(5.8571804183181\times 10^{-2}\)\\
8&\(4.15552621898715\times 10^{-2}\)&\(4.23286430116071\times 10^{-2}\)\\
9&\(3.16509341842258\times 10^{-2}\)&\(3.20885498792370\times 10^{-2}\)\\
10&\(2.49263586496073\times 10^{-2}\)&\(2.51925841277600\times 10^{-2}\)\\
12&\(1.66246017064658\times 10^{-2}\)&\(1.67396850792489\times 10^{-2}\)\\
14&\(1.18801077328929\times 10^{-2}\)&\(1.19376803778606\times 10^{-2}\)\\
16&\(8.91343515555617\times 10^{-3}\)&\(8.94533890457622\times 10^{-3}\)\\
18&\(6.93472364128979\times 10^{-3}\)&\(6.95379534903761\times 10^{-3}\)\\
20&\(5.54905996206358\times 10^{-3}\)&\(5.56114762804262\times 10^{-3}\)\\
30&\(2.37962509465255\times 10^{-3}\)&\(2.38177780442984\times 10^{-3}\)\\
40&\(1.31536713121881\times 10^{-3}\)&\(1.31601370250459\times 10^{-3}\)\\
50&\(8.33160391138597\times 10^{-4}\)&\(8.33417027210234\times 10^{-4}\)\\
60&\(5.74629392014362\times 10^{-4}\)&\(5.74750573327324\times 10^{-4}\)\\
70&\(4.20123225270203\times 10^{-4}\)&\(4.20187653749219\times 10^{-4}\)\\
80&\(3.20486589318390\times 10^{-4}\)&\(3.20523928418601\times 10^{-4}\)\\
90&\(2.52508907696752\times 10^{-4}\)&\(2.52532012109719\times 10^{-4}\)\\
100&\(2.04070927223777\times 10^{-4}\)&\(2.04085978386760\times 10^{-4}\)\\
500&\(8.03211169423443\times 10^{-6}\)&\(8.03213455900903\times 10^{-6}\)\\
1000&\(2.00400696809141\times 10^{-6}\)&\(2.00400838779413\times 10^{-6}\)\\
5000&\(8.00320111329434\times 10^{-8}\)&\(8.00320133925429\times 10^{-8}\)\\
\end{tabular}
\end{ruledtabular}
\end{table}

\end{widetext}


\begin{acknowledgments}
JET acknowledges the support of several people in the completion of this work, including Abhay Shah, Adam Pound, Niels Warburton, and Seth Hopper.
The authors would like to thank the referee for providing several insightful comments and suggestions to this work.
Plots and figures in this text were generated with the use of the \textsc{numpy} \cite{numpy} and \textsc{matplotlib} \cite{matplotlib} packages.
The work of JET and BFW was supported in part by NSF Grants PHY 1314529 and PHY
1607323 to the University of Florida. Support from the CNRS through the
IAP, where part of this work was carried out, is also acknowledged, along
with support from the French state funds managed by the ANR within the
Investissements d'Avenir programme under Grant No. ANR-11-IDEX-0004-02.
\end{acknowledgments}

\begin{appendices}


\section{Tensor Harmonic Basis \label{tensorbasis}}

\subsection{Vector and Tensor Harmonics}
We review the pure-spin tensor-harmonic basis introduced in \cite{ThorneRMP52} which is used with the A--K notation in Eq.~\eqref{eq:hina-k}. The scalar spherical harmonics are defined as eigenfunctions of the spherical Laplacian and given by,
\beq
Y_{\ell m}(\theta,\phi)=\sqrt{\frac{(2\ell+1)}{4\pi}\frac{(\ell-m)!}{(\ell+m)!}}P_\ell^m(\cos\theta)e^{im\phi},
\eeq
where \(P_\ell^m(\cos\theta)\) is the associated Legendre polynomial. In constructing the vector- and tensor-harmonics, we require the two vector fields \(v_a\) and \(n_a\), introduced in Sec.~\ref{retsol}, along with the Schwarzschild metric on the two-sphere,
\beq
\sigma_{ab}=g_{ab}+fv_av_b-f^{-1}n_an_b.
\eeq
The vector-harmonics are now defined as,
\begin{align}
Y^{E,\ell m}_a&=r\nabla_aY_{\ell m},\\
Y^{B,\ell m}_a&=r\varepsilon_{ab}{}^{c}n^b\nabla_cY_{\ell m},\\
Y^{R,\ell m}_a&=n_aY_{\ell m},
\end{align}
where \(\varepsilon_{abc}\) is the spatial Levi-Civita tensor with \(v^a\varepsilon_{abc}=0\) and \(\varepsilon_{r\theta\phi}=r^2\sin\theta\). The tensor-harmonics are further defined as,
\begin{align}
T^{T0,\ell m}_{ab}&=\sigma_{ab}Y_{\ell m},\\
T^{L0,\ell m}_{ab}&=n_an_bY_{\ell m},\\
T^{E1,\ell m}_{ab}&=rn_{(a}\nabla_{b)}Y_{\ell m},\\
T^{B1,\ell m}_{ab}&=rn_{(a}\varepsilon_{b)c}{}^{d}n^c\nabla_dY_{\ell m},\\
T^{E2,\ell m}_{ab}&=r^2(\sigma_{a}{}^{c}\sigma_{b}{}^{d}-\frac12\sigma_{ab}\sigma^{cd})\nabla_c\nabla_dY_{\ell m},\\
T^{B2,\ell m}_{ab}&=r^2\sigma_{(a}{}^{c}\varepsilon_{b)e}{}^{d}n^e\nabla_c\nabla_dY_{\ell m}.
\end{align}
Finally we list the conventions used for finding the A--K projections introduced in \cite{TCW}. For an arbitrary smooth tensor field \(X_{ab}\), the A--K terms are found via,
\pagebreak
\begin{align}
\label{Xa}X_\tA&=f^2\int v^av^bY_{\ell m}^*X_{ab}\,\rmd\Omega,\\
X_\tB&=-\frac{f}{\ell(\ell+1)}\int v^aY^{b\,*}_EX_{ab}\,\rmd\Omega,\\
X_\tC&=-\frac{f}{\ell(\ell+1)}\int v^aY^{b\,*}_BX_{ab}\,\rmd\Omega,\\
X_\tD&=-\int v^aY_R^{b\,*}X_{ab}\,\rmd\Omega,\\
X_\tE&=\frac12\int T^{ab\,*}_{T0}X_{ab}\,\rmd\Omega,\\
X_\tF&=\frac{2(\ell-2)!}{(\ell+2)!}\int T^{ab\,*}_{E2}X_{ab}\,\rmd\Omega,\\
X_\tG&=\frac{2(\ell-2)!}{(\ell+2)!}\int T^{ab\,*}_{B2}X_{ab}\,\rmd\Omega,\\
X_\tH&=\frac{(\ell-1)!}{(\ell+1)!}f^{-1}\int T^{ab\,*}_{E1}X_{ab}\,\rmd\Omega,\\
X_\tJ&=\frac{(\ell-1)!}{(\ell+1)!}f^{-1}\int T^{ab\,*}_{B1}X_{ab}\,\rmd\Omega,\\
\label{Xk}X_\tK&=f^{-2}\int T^{ab\,*}_{L0}X_{ab}\,\rmd\Omega,
\end{align}
where \(*\) denotes complex conjugation and \linebreak\(\rmd\Omega=\sin\theta\rmd\theta\rmd\phi\).

\subsection{Rotations and \(m\)-Sums \label{rotations}}
To perform the rotation between the two-sphere angles \((\Theta,\Phi)\) and the original Schwarzschild angles (\(\theta,\phi\)) in Sec.~\ref{singfieldconst}, we use the Wigner-D matrices \(D^\ell_{m,m'}\),
\beq
D^\ell_{m,s}(\alpha,\beta,\gamma)=(-1)^s\sqrt{\frac{4\pi}{2\ell+1}}{}_{-s}Y_{\ell m}^*(-\beta,\alpha)e^{-is\gamma},
\eeq
written here for the Euler angles \(\alpha\), \(\beta\), and \(\gamma\) chosen in \cite{WardellPRD92} and in terms of spin-weighted spherical harmonics \({}_sY_{\ell m}(\theta,\phi)\) using the conventions of \textsc{Mathematica}~\cite{Mathematica}. The spin-weighted spherical harmonics may be constructed from the scalar spherical harmonics \cite{Shah2016}; for \(s=0\), the spin-weighted and scalar spherical harmonics are related via the identification
\beq
{}_0Y_{\ell m}(\theta,\phi)=Y_{\ell m}(\theta,\phi).
\eeq
To construct a spin-weighted harmonic \({}_sY_{\ell m}\) with spin-weight \(s\), raising and lowering operators are defined, respectively,
\begin{align}
\eth_s&=-\frac{\partial_\theta+i\csc\theta\partial_\phi-s\cot\theta}{\sqrt{(l-s)(l+s+1)}},\\
\bar{\eth}_s&=\frac{\partial_\theta-i\csc\theta\partial_\phi+s\cot\theta}{\sqrt{(l+s)(l-s+1)}},
\end{align}
such that any spin-weight \(s\) harmonic is achieved by repeated application of a raising or lowering operator on \(Y_{\ell m}\), e.g.,
\begin{align}
{}_1Y_{\ell m}(\theta,\phi)&=\eth_0\,Y_{\ell m}(\theta,\phi),\\
{}_{-2}Y_{\ell m}(\theta,\phi)&=\bar{\eth}_{-1}\bar{\eth}_0\,Y_{\ell m}(\theta,\phi).
\end{align}
It is clear that any spin-weighted spherical harmonic may be written as a combination of scalar spherical harmonics and their angular derivatives.

We noted in Sec.~\ref{THRP} of the difficulty involved in performing the \(m\)-sum of the singular field analytically, as the sum is performed in the original, unrotated Schwarzschild coordinates. After reconstructing the singular field, each component of \(h_{ab}^{\text{RW},\tS,\ell m}\) contains terms with the following two forms produced by the rotation in Eq.~\eqref{ASfullrotation}:
\begin{align}
\nonumber\text{for even parity,}&\\
\label{wigDplus}D^\ell_{\,m,m'}\left(\pi,\frac\pi2,\right.&\!\!\left.\frac\pi2\right)+(-1)^{m'}D^\ell_{\,m,-m'}\left(\pi,\frac\pi2,\frac\pi2\right),\\
\nonumber\text{for odd-parity,}&\\
\label{wigDminus}D^\ell_{\,m,m'}\left(\pi,\frac\pi2,\right.&\!\!\left.\frac\pi2\right)-(-1)^{m'}D^\ell_{\,m,-m'}\left(\pi,\frac\pi2,\frac\pi2\right).
\end{align}
These expressions result from combining \(\pm m'\) values together and simplifying using the complex conjugation of the A--K variables, e.g., \(\tA^{\ell,-m'}=(-1)^{m'}\tA^{\ell,m' *}\). We write these combinations of Wigner-D matrices in terms of spherical harmonics,
\begin{widetext}
\begin{align}
\label{wigDplustoylm}D^\ell_{\,m,m'}\left(\pi,\frac\pi2,\frac\pi2\right)+(-1)^{m'}D^\ell_{\,m,-m'}\left(\pi,\frac\pi2,\frac\pi2\right)&=\sqrt{\frac{4\pi}{2\ell+1}}\,a^+_{\ell m m'}\,Y^*_{\ell m}(\frac{\pi}{2},0),\\
\label{wigDminustoylm}D^\ell_{\,m,m'}\left(\pi,\frac\pi2,\frac\pi2\right)-(-1)^{m'}D^\ell_{\,m,-m'}\left(\pi,\frac\pi2,\frac\pi2\right)&=\sqrt{\frac{4\pi}{2\ell+1}}\,a^-_{\ell mm'}\,\partial_\theta Y^*_{\ell m}(\frac{\pi}{2},0),
\end{align}
with coefficients,
\beq
a^+_{\ell mm'}=\sqrt{\frac{(\ell-m')!}{(\ell+m')!}}\left\{
\begin{tabular}{c c}
1& for \(m'=0\),\\
\(2im\) & for \(m'=1\),\\
\(2[\ell(\ell+1)-2m^2]\) & for \(m'=2\),\\
\end{tabular}
\right.
\eeq
\beq
a^-_{\ell mm'}=\sqrt{\frac{(\ell-m')!}{(\ell+m')!}}\left\{
\begin{tabular}{c c}
0 & for \(m'=0\),\\
\(2i\) & for \(m'=1\),\\
\(-4m\) & for \(m'=2\),\\
\end{tabular}
\right.
\eeq
 reducing the necessary sums over \(m\)-modes to be proportional to either,
\begin{align}
\label{ylmsum}&\sum_{m=-\ell}^\ell m^N\left|Y_{\ell m}\left(\pi/2,0\right)\right|^2,\\
\nonumber\text{or},&\\
\label{dylmsum}&\sum_{m=-\ell}^\ell m^N\left|\partial_\theta Y_{\ell m}\left(\pi/2,0\right)\right|^2.
\end{align}
The sums in Eqs.~\eqref{ylmsum} and \eqref{dylmsum} are calculated analytically by Nakano \etal \cite{NakanoPRD68}, who evaluate them by repeated differentiation of two generating functions,
\begin{align}
\sum_{m=-\ell}^\ell m^N\left|Y_{\ell m}\left(\pi/2,0\right)\right|^2&=\lim_{z\rightarrow 0}\frac{\rmd^N}{\rmd z^N}\left[\frac{2\ell+1}{4\pi}e^{\ell z}{}_2F_1\left(\frac12,-\ell,1,1-e^{-2z}\right)\right],\\
\nonumber\sum_{m=-\ell}^\ell m^N\left|\partial_\theta Y_{\ell m}\left(\pi/2,0\right)\right|^2&=\lim_{z\rightarrow 0}\frac{\rmd^N}{\rmd z^N}\left[
\frac{2\ell+1}{4\pi}e^{(\ell-1)z}\frac{\Gamma(\ell+1/2)\Gamma(3/2)}{\Gamma(\ell)}\right.\\
&\qquad\qquad\qquad\left.\times\,{}_2F_1\left(\frac32,-\ell+1,-\ell+\frac12,e^{-2z}\right)
\right].
\end{align}
\end{widetext}


\section{Source Terms in A--K \label{sourceinAK}}
We first list the source terms for the master functions in Eq.~\eqref{waveeq},
\begin{align}
\label{sourceodd}S_\tW&=f\left(r^2\partial_rE_\tC+rE_\tC+r^2\partial_tE_\tJ\right),\\
\nonumber S_\tZ&=\frac{-rf}{2\kappa}\left[-\frac{r[\lambda(\lambda-2)r^2+2Mr(7\lambda-18)+96M^2]}{2rf\kappa}E_\tA\right.\\
\nonumber&\qquad\qquad+r^2\partial_rE_\tA+r^2\partial_tE_\tD\\
\label{sourceeven}&\;\;\;\;\;\;\;\;\;\;\;\;\;\;\left.+(\lambda+2)\left(rfE_\tH+\frac{rf}{2}E_\tK-\frac{\kappa}{2}E_\tF\right)\right].
\end{align}

The source terms are constructed from projections of the stress-energy tensor onto the tensor-harmonic basis, Eq.~\eqref{sourceprojections} and Eqs.~\eqref{Xa}-\eqref{Xk}. When evaluated for a circular orbit, the non-vanishing source terms are:
\begin{align}
\nonumber\ell\ge0,&\\
E^{\ell m}_\text{A}&=-16\pi \frac{\mu f_0\mathcal{E}}{r_0^2}\delta\left(r-r_0\right)Y^{*}_{\ell m}\left(\frac\pi2,\Omega t\right),\label{eq:stressA}\\
E^{\ell m}_\text{E}&=-8\pi\frac{\mu \Omega \mathcal{L}}{r_0^2}\delta\left(r-r_0\right)Y^{*}_{\ell m}\left(\frac\pi2,\Omega t\right),\\
\nonumber\ell\ge1,&\\
E^{\ell m}_\text{B}&=\frac{16\pi im}{\ell(\ell+1)}\frac{\mu f_0\mathcal{L}}{r_0^3}\delta\left(r-r_0\right)Y^{*}_{\ell m}\left(\frac\pi2,\Omega t\right),\\
E^{\ell m}_\text{C}&=\frac{-16\pi}{\ell(\ell+1)}\frac{\mu f_0\mathcal{L}}{r_0^3}\delta\left(r-r_0\right)\partial_\theta Y^{*}_{\ell m}(\frac\pi2,\Omega t),\label{eq:stressC}
\end{align}
\pagebreak
\begin{align}
\nonumber\ell\ge2,&\\
\nonumber E^{\ell m}_\text{F}&=-16\pi\frac{(\ell-2)!}{(\ell+2)!}\frac{\mu\Omega\mathcal{L}}{r_0^2}\delta\left(r-r_0\right)\\
&\quad\times\left[\ell(\ell+1)-2m^2\right]Y^{*}_{\ell m}\left(\frac\pi2,\Omega t\right).
\end{align}

\section{A--K and Barack-Lousto-Sago Decompositions}
\label{BLSbasis}
For convenience, we list the A--K variables of the metric perturbation in terms of the BLS basis of Barack and Lousto \cite{BarackPRD66} and Barack and Sago \cite{BarackPRD75},
\begin{align*}
\tA&=\frac1 {2r}\left(\bar{h}^{(1)}+f\bar{h}^{(6)}\right),\\
\tB&=-\frac1{2r}\frac1{\ell(\ell+1)}\hb{4},\\
\tC&=\frac1{2r}\frac1{\ell(\ell+1)}\hb{8},\\
\tD&=-\frac1{2rf}\hb{2},\\
\tE&=\frac1{2r}\hb{3},\\
\tF&=\frac{(\ell-2)!}{(\ell+2)!}\frac{\hb{7}}{r},\\
\tG&=-\frac{(\ell-2)!}{(\ell+2)!}\frac{\hb{10}}{r},\\
\tH&=\frac1{2rf}\frac1{\ell(\ell+1)}\hb{5},\\
\tJ&=-\frac1{2rf}\frac1{\ell(\ell+1)}\hb{9},\\
\tK&=\frac1{2rf^2}\left(\hb{1}-f\hb{6}\right).
\end{align*}

\begin{widetext}

\section{Additional Force Regularization Parameters}
\label{forceregparam}

In this section, we present the results of the singular gauge transformation, which contributes to the sub-leading self-force regularization parameters for the self-force in the EZ and RW gauges. We also display the \(D^r\) regularization parameter for both the EZ and RW gauges.

\subsection{EZ Gauge}
The \(\ell\)-modes of the singular gauge vector from the Lorenz to EZ gauge are found to be,
\begin{align}
\nonumber\xi^{r,\ell}_\text{EZ,S}&=\frac{2\mu}{\pi  (r_0-3 M)^{1/2} (r_0-2 M)^{1/2}}\left\{ 2( r_0-2 M)\hat{\mathcal{E}}-(r_0-3M)\hat{\mathcal{K} }\right\}\\
\nonumber&\quad+\bigg[\frac{16\mu}{10395\pi r_0(r_0-3M)^{3/2}(r_0-2M)^{1/2}}\left\{ (2 M-r_0) \left(4805 M^2-14843 M r_0+4896 r_0^2\right)\hat{\mathcal{E}}\right.\\
&\quad\quad\quad\left.+ (r_0-3M) \left(6820 M^2-14255 M r_0+4896  r_0^2\right)\hat{\mathcal{K}}\right\}\bigg]\delta_{\ell 1}.
\end{align}
The \(D^r_\text{EZ}\) regularization parameter is given by,
\begin{align}
\nonumber D^r_\text{EZ}&=\frac{16M(r_0-2M)^{1/2}}{3465\pi r_0^4(r_0-3M)^{5/2}}\left\{ (r_0-2M)\left(4805 M^2-14843 M r_0+4896 r_0^2\right)\hat{\mathcal{E}}\right.\\
\label{DEZ}&\quad\quad\quad\left.-(r_0-3M) \left(6820 M^2-14255 M r_0+4896 r_0^2\right)\hat{\mathcal{K}}\right\}
\end{align}

\pagebreak
\subsection{RW Gauge}
The \(\ell\)-modes of the singular gauge vector from the Lorenz to RW gauge are found to be,
\begin{align}
\nonumber\xi^{r,\ell}_\text{RW,S}&=
-\bigg[\frac{\mu(r_0-2M)^{1/2}}{945\pi r_0^2(r_0-3M)^{3/2}}\left\{ \left(21824 M^3-56310 M^2 r_0+32677 M r_0^2-4269 r_0^3\right)\hat{\mathcal{E}}\right.\\
\nonumber&\quad\quad\quad\left.+ (r_0-3M) \left(15488 M^2-14416 M r_0+2379   r_0^2\right)\hat{\mathcal{K}}\right\}\bigg]\delta_{\ell 0}\\
\nonumber&\quad+\bigg[\frac{\mu(r_0-2M)^{1/2}}{10395\pi r_0^2(r_0-3M)^{3/2}}\left\{ \left(307520 M^3-566142 M^2 r_0+311041 M r_0^2-47145 r_0^3\right)\hat{\mathcal{E}}\right.\\
&\quad\quad\quad\left.+(r_0-3M) \left(218240 M^2-174736 M r_0+67935   r_0^2\right)\hat{\mathcal{K}}\right\}\bigg]\delta_{\ell 1}.
\end{align}
The \(D^r_\text{RW}\) regularization parameter is given by,
\begin{align}
\nonumber D^r_\text{RW}&=-\frac{\mu^22M(r_0-2M)^{3/2}}{3465\pi r_0^5(r_0-3M)^{5/2}}\left\{(33728M^3+26634M^2r_0-24203Mr_0^2-93r_0^3)\hat{\mathcal{E}}\right.\\
\label{DRW}&\quad\quad+\left.(r_0-3M)(23963M^2-8080Mr_0+20883r_0^2)\hat{\mathcal{K}}\right\}.
\end{align}
\pagebreak
\end{widetext}


\section{Local Gauge Transformation from Lorenz to EZ \label{LorenzToEZ}}

The EZ  gauge condition \cite{TCW} is typically reported as an algebraic condition on various tensor-harmonic mode components of the metric perturbation. This form of the gauge condition assumes a global decomposition of the metric perturbation into tensor-harmonic modes, and the gauge condition is applied mode-by-mode; such a decomposition is not locally defined and fails to describe the local behavior of a gauge transformation to the EZ gauge. We wish to study this \textit{local} behavior of the gauge transformation from the Lorenz gauge to the EZ gauge, and as such must look at the more general form of the EZ gauge condition, namely:
\begin{align}
\label{gc1}h_{\theta\theta}^{\text{EZ}}&=0,\\
\label{gc2}h_{\phi\phi}^{\text{EZ}}&=0,\\
\label{gc3}h_{\theta\phi}^{\text{EZ}}&=0,\\
\label{gc4}\sin\theta\left(\sin\theta\, h_{t\theta}^{\text{EZ}}\right)_{,\theta}+h_{t\phi,\phi}^{\text{EZ}}&=0.
\end{align}
Gauge conditions \eqref{gc1}--\eqref{gc3} state that the components of the metric perturbation on the two-sphere are set to zero (in A--K, the E, F, and G terms), and the gauge condition \eqref{gc4} is used to eliminate one even-parity vector piece of the metric perturbation (the B term). This form of the EZ gauge condition is well-suited for a local investigation of the gauge vector, and is also satisfied automatically by the \(l=0,1\) Zerilli gauge monopole and dipole.

The gauge transformation from the Lorenz gauge to the EZ gauge is generated by the vector \(\xi^a\). To first order in the gauge vector, this transformation takes the form,
\begin{align}
h_{ab}^{\text{EZ}}=h_{ab}^{\text{L}}-2\nabla_{(a}\xi_{b)}.
\end{align}
When substituted into the gauge conditions \eqref{gc1}--\eqref{gc4}, the gauge vector must satisfy the following equations,
\begin{align}
\label{gc1E}h_{\theta\theta}^\text{L}&=2\,\xi_{\theta,\theta}+2(r-2M)\xi_r,\\
\label{gc2E}h_{\phi\phi}^\text{L}&=2\,\xi_{\phi,\phi}+2(r-2M)\sin^2\theta\,\xi_r\\
\nonumber&\qquad+2\sin\theta\cos\theta\xi_\theta,\\
\label{gc3E}h_{\theta\phi}^\text{L}&=\xi_{\theta,\phi}+\sin^2\theta\,(\sin^{-2}\theta\,\xi_\phi)_{,\theta},\\
\label{gc4E}\sin\theta&\,(\sin\theta\, h_{t\theta}^\text{L})_{,\theta}+h_{t\phi,\phi}^{\text{L}}\\
\nonumber&=\sin\theta\,(\sin\theta\, \xi_{t,\theta})_{,\theta}+\xi_{t,\phi\phi}\\
\nonumber&\qquad+\sin\theta\,(\sin\theta\, \dot{\xi}_{\theta})_{,\theta}+\dot{\xi}_{\phi,\phi},
\end{align}
where an overdot represents a time derivative. To analyze these equations, we follow the framework laid out by Barack and Ori (BO) \cite{BarackPRD64}. As the gauge equations do not contain any radial derivatives, we choose to work on a constant \(r=r_0\) hypersurface. Furthermore, we may recover Eqs. (38) of BO (up to a sign convention) by combining Eqs.~\eqref{gc1E} and \eqref{gc2E} as defined for \(h_\text{ang}\equiv(h_{\theta\theta}-\sin^{-2}\theta\, h_{\phi\phi})/2\), which eliminates \(\xi_r\):
\begin{align}
\label{gauge_ang}\sin\theta\,(\sin^{-1}\theta\,\xi_{\theta})_{,\theta}-\sin^{-2}\theta\,\xi_{\phi,\phi}&=h_\text{ang}^\text{L}.
\end{align}

The resulting equations naturally separate into conditions on the angular components \(\xi_\theta\) and \(\xi_\phi\), Eqs.~\eqref{gc3E} and \eqref{gauge_ang}, and the time component \(\xi_t\), Eq.~\eqref{gc4E}.


\subsection{Solving for \(\xi_\theta\) and \(\xi_\phi\)}
\label{sec:solve_for_xiBO}

We reproduce the results of BO here for completeness. To simplify the work involved, BO observe that Eqs.~\eqref{gc3E} and \eqref{gauge_ang} do not involve time derivatives, so we may further restrict our analysis to the surface \((t=0,r=r_0)\).

The local Lorenz gauge singular field may be written for a perturbing mass \(\mu\) as in Eq.~\eqref{lorenzsing},
\begin{equation}
h_{ab}^\text{L}=\frac{2\mu}{s}(g_{ab}+2u_au_b),
\end{equation}
where \(u_a\) is the four-velocity of the particle and \(s\) is the spatial geodesic displacement away from the worldline along the surface. We may then re-write Eqs,~\eqref{gc3E} and \eqref{gauge_ang}, introducing the singular fields as a source term on the RHS:
\begin{align}
\label{gauge_ang_source}\sin\theta\,(\sin^{-1}\theta\,\xi_{\theta})_{,\theta}-\sin^{-2}\theta\,\xi_{\phi,\phi}&=-\frac{2\mu\mathcal{L}^2}{s},\\
\label{gc3E_source}\xi_{\theta,\phi}+\sin^2\theta\,(\sin^{-2}\theta\,\xi_\phi)_{,\theta}&=0.
\end{align}
Following BO, we now perform a change of coordinates on the two-sphere to be cartesian-like: \(y=r_0\sin\theta\sin\varphi\), \(z=r_0\cos\theta\). To see how this coordinate transformation affects \(s\), we use the definition of the space-like interval along the submanifold spanned by \(y\) and \(z\),
\begin{align*}
s&=\sqrt{(g_{ab}+u_au_b-n_an_b)x^ax^b}\\
&=\sqrt{\left(1+\frac{\mathcal{L}^2}{r_0^2}\right)y^2+z^2}+o(y)+o(z).
\end{align*}
In their paper, BO define the quantity
\[
(1-v^2)^{-1}=\left(1+\frac{\mathcal{L}^2}{r_0^2}\right),
\]
where \(0<v<1\) is the local boost velocity. After expanding out Eqs.~\eqref{gauge_ang_source} and \eqref{gc3E_source}, we find to leading order,
\begin{align}
\label{yminusz}\xi_{z,z}-\xi_{y,y}&=-\frac{2\mu\mathcal{L}^2}{r_0^2\sqrt{(1-v^2)^{-1}y^2+z^2}},\\
\label{yplusz}\xi_{z,y}+\xi_{y,z}&=0.
\end{align}
Eq.~\eqref{yplusz} implies that both \(\xi_y\) and \(\xi_z\) can be found by differentiating a scalar potential \(\Phi\),
\begin{equation}
\label{gaugefrompotential}\xi_z=\Phi_{,z},\;\;\;\xi_y=-\Phi_{,y},
\end{equation}
which must satisfy Poisson's equation,
\begin{equation}
\label{potentialpoisson}\Phi_{,zz}+\Phi_{,yy}=-\frac{2\mu\mathcal{L}^2}{r_0^2\sqrt{(1-v^2)^{-1}y^2+z^2}}.
\end{equation}
At this point, we transform coordinates again, changing the local cartesian coordinates to the polar coordinates \(y=\rho\cos\alpha\), \(z=\rho\sin\alpha\), and re-express Eq.~\eqref{potentialpoisson},
\begin{equation}
\label{potentialpoissonpolar}\frac1\rho\left(\rho\Phi_{,\rho}\right)_{,\rho}+\frac1{\rho^2}\Phi_{,\alpha\alpha}=\frac{a}{\rho}\frac{(1-v^2)^{1/2}}{\sqrt{1-v^2\sin^2\alpha}},
\end{equation}
with \(a=-2\mu\mathcal{L}^2/r_0^2\). If we suppose an Ansatz for the solution which is decomposed into Fourier modes \(e^{in\alpha}\), we find that the general form of the potential is,
\begin{equation}
\Phi(\rho,\alpha)=c\alpha+\sum_{n=-\infty}^{\infty}e^{in\alpha}\Phi_n(\rho).
\end{equation}
The term \(c\alpha\) in this general solution exists because the potential \(\Phi\) need not be single-valued, due to the presence of the singularity at \(y=z=0\), but its \(\rho\)- and \(\alpha\)-derivatives must be single-valued. After substitution into Eq.~\eqref{potentialpoissonpolar}, the Fourier modes \(\Phi_n\) obey the equation,
\begin{equation}
\label{potentialmodes}\frac1\rho\left(\rho\Phi_{n,\rho}\right)_{,\rho}-\frac{n^2}{\rho^2}\Phi_n=\frac{a}{\rho}f_n,
\end{equation}
with coefficients,
\begin{equation}
f_n=\frac{\sqrt{1-v^2}}{2\pi}\int_0^{2\pi}\frac{e^{-in\alpha}}{\sqrt{1-v^2\sin^2\alpha}}d\alpha.
\end{equation}
These coefficients vanish for odd \(n\), and are generally non-vanishing for even \(n\). In particular,
\[
f_0=\frac{2\sqrt{1-v^2}}{\pi}\mathcal{K}(v^2),
\]
written in terms of the complete elliptic integral of the first kind, \(\mathcal{K}\), is bounded from below away from zero. (This \(f_0\) is not to be confused with \(f(r_0)\) used in the body of this paper.) BO next construct the general solution to Eq.~\eqref{potentialmodes},
\begin{equation}
\Phi_n=\left\{
\begin{tabular}{l c r}
\(b_0\rho+\gamma_0+\beta_0\ln\rho\) & for & \(n=0\),\\
\(b_n\rho+\gamma_n\rho^{|n|}+\beta_n\rho^{-|n|}\) & for & \(n\ne 0,\)\\
\end{tabular}\right.
\end{equation}
with the arbitrary constants \(\gamma_n\) and \(\beta_n\) arising from the homogeneous solutions and the constants \(b_n\) determined by the particular solution,
\begin{equation}
\label{BOfouriermodes}b_n=\left\{
\begin{tabular}{l l}
\(af_n/(1-n^2)\) & for even \(n\),\\
0 & for odd \(n\).\\
\end{tabular}\right.
\end{equation}
With the general solution determined, the task is now to find the \textit{most regular} behavior of the gauge vector as we approach the worldline (in this case, as \(\rho\rightarrow 0\)). BO find the most regular solution to be one which sets \(\beta_n=0\) for all values of \(n\), along with \(c=0\). They then write the final solution in a compact form,
\begin{equation}
\Phi(\rho,\alpha)=\gamma_0+\rho H(\alpha)+O(\rho^2),
\end{equation}
with
\[
H(\alpha)=\gamma_1e^{i\alpha}+\gamma_{-1}e^{-i\alpha}+\sum_{n=-\infty}^{\infty}b_ne^{in\alpha}.
\]
Finally, the components of the gauge vector are recovered by differentiating the potential, \textit{a la} Eqs.~\eqref{gaugefrompotential},
\begin{align}
\label{xi_y_sol}\xi_y&=-H\cos\alpha+H_{,\alpha}\sin\alpha+O(\rho),\\
\label{xi_z_sol}\xi_z&=H\sin\alpha+H_{,\alpha}\cos\alpha+O(\rho).
\end{align}
For the components of the gauge vector to be continuous at the particle, they must be independent of \(\alpha\), otherwise the \(\rho\rightarrow0\) limit takes an indefinite value. BO find, though, that the first derivative of the gauge vector components, \(\xi_{y,\alpha}\) and \(\xi_{z,\alpha}\), do not vanish at the particle, implying directional dependence to their values and the presence of a jump discontinuities. They stress that, while \(\xi_y\) and \(\xi_z\) are discontinuous at the particle, they remain \textit{bounded} in the limit, thereby still satisfying the sufficiently regular criteria.


\subsection{Solving for \(\xi_t\)}
\label{sec:localxit}
 We now look to solve Eq.~\eqref{gc4E} for \(\xi_t\). Following the lead of Pound, Merlin, and Barack~\cite{PoundPRD89}, we now find a solution for \(\xi_a\) which is well-behaved as a function of time and satisfies (SR3). As such, the time derivatives in Eq.~\eqref{gc4E} are subdominant to the spatial derivatives when looking at the most singular behavior, and are ignored, reducing Eq.~\eqref{gc4E} to
\begin{equation}
\label{gc4Ereduced}\sin\theta\,(\sin\theta\, \xi_{t,\theta})_{,\theta}+\xi_{t,\phi\phi}=\sin\theta\,(\sin\theta\, h_{t\theta}^\text{L})_{,\theta}+h_{t\phi,\phi}^{\text{L}}.
\end{equation}
 For a particle traveling along a circular geodesic of Schwarzschild spacetime, the RHS of Eq.~\eqref{gc4Ereduced} becomes,
\begin{equation}
\label{gc4Ereduced_source}\sin\theta\,(\sin\theta\, h_{t\theta}^\text{L})_{,\theta}+h_{t\phi,\phi}^{\text{L}}=\left[\frac{-4\mu\mathcal{E}\mathcal{L}}{s}\right]_{,\phi}.
\end{equation}
We again introduce the locally cartesian coordinates \((y,z)\) on the surface \((t=0,r=r_0)\), and expand Eq.~\eqref{gc4Ereduced_source}, keeping only the leading terms,
\begin{equation}
\xi_{t,yy}+\xi_{t,zz}=\frac{4\mu  \mathcal{E}\mathcal{L}(1-v^2)^{-1}  y
  }{r_0 \left[(1-v^2)^{-1} y^2+z^2\right]^{3/2}}.
\end{equation}
The LHS is simply the flat-space Laplacian acting on \(\xi_t\). When transformed to the polar coordinates used in Sec.~\ref{sec:solve_for_xiBO}, the equation becomes equivalent to Eq.~\eqref{potentialpoissonpolar} with a different source term,
\begin{equation}
\label{potentialpoisson_xit}\frac1\rho\left(\rho\,\xi_{t,\rho}\right)_{,\rho}+\frac1{\rho^2}\xi_{t,\alpha\alpha}=\frac{c}{\rho^2}\frac{(1-v^2)^{1/2}\cos\alpha}{[1-v^2\sin^2\alpha]^{3/2}},
\end{equation}
with \(c=4\mu\mathcal{E}\mathcal{L}/r_0\). When decomposed into Fourier modes, \(\xi_t\) is expressed as,
\[
\xi_t=\sum_{n=-\infty}^{\infty}e^{in\alpha}\xi_t^n(\rho),
\]
satisfying,
\begin{equation}
\frac1\rho\left(\rho\,\xi^n_{t,\rho}\right)_{,\rho}-\frac{n^2}{\rho^2}\xi^n_t=\frac{c}{\rho^2}d_n.
\end{equation}
The Fourier modes of the source now have different characteristics,
\begin{equation}
d_n=\frac{\sqrt{1-v^2}}{2\pi}\int_0^{2\pi}\frac{e^{-in\alpha}\cos\alpha}{(1-v^2\sin^2\alpha)^{3/2}}d\alpha,
\end{equation}
which vanish for all \textit{even} values of \(n\) and are generally non-vanishing for odd values of \(n\), yielding the opposite behavior of the coefficients \(f_n\). We again construct the most general solution to Eq.~\eqref{potentialpoisson_xit},
\begin{equation}
\xi^n_t=\left\{
\begin{tabular}{l c r}
\(\gamma_0+\beta_0\ln\rho\) & for & \(n=0\),\\
\(q_n+\gamma_n\rho^{|n|}+\beta_n\rho^{-|n|}\) & for & \(n\ne 0.\)\\
\end{tabular}\right.
\end{equation}
The constants \(\gamma_n\) and \(\beta_n\) are again arbitrary, and \(q_n\) is defined as,
\begin{equation}
\label{fourier_coeff}q_n=\left\{
\begin{tabular}{l l}
0 & for even \(n\),\\
\(-c\,d_n/n^2\) & for odd \(n\).\\
\end{tabular}\right.
\end{equation}
It is clear that the most regular solution may be obtained by setting \(\beta_n=0\) for all values of \(n\), but we note that \(\beta_0\ne0\) is still allowed by the regularity condition of (SR1). Finally we define the function \(G(\alpha)\) in a similar way to \(H(\alpha)\),
\begin{equation}
G(\alpha)=\gamma_1e^{i\alpha}+\gamma_{-1}e^{-i\alpha},
\end{equation}
such that the full gauge vector component \(\xi_t\) is recovered,
\begin{equation}
\label{xi_t}\xi_t=\gamma_0+\sum_{n=-\infty}^{\infty}q_ne^{in\alpha}+\rho \,G(\alpha)+O(\rho^2).
\end{equation}

The sum in Eq.~\eqref{xi_t} converges for any value of \(\alpha\in[0,2\pi)\), and thus \(\xi_t\) is well-behaved in the \(\rho\rightarrow0\) limit yet still dependent on \(\alpha\), indicating a jump discontinuity.


\subsection{Solving for \(\xi_r\)} Finally, we solve for the radial component of the gauge vector, \(\xi_r\), by combining Eqs.~\eqref{gc1E} and \eqref{gc2E} as in Eq.~\eqref{gauge_ang} but by adding the equations instead of subtracting. Returning once again to the cartesian coordinates for \(\xi_\theta\) and \(\xi_\phi\),
\begin{equation}
\xi_r=\frac{r_0^2}{2(r_0-2M)}\left(\frac{\mu(1-v^2)^{-1}}{s}-\xi_{y,y}-\xi_{z,z}\right).
\end{equation}
This equation seems to indicate that \(\xi_r\) (restricted to the two-sphere intersecting the worldline of the particle) diverges as \(1/s\) as one approaches the particle. Such a divergence is too singular to fall within the class of sufficiently regular gauge transformations, for it does not satisfy (SR2). On closer inspection, in the local polar coordinates and using Eqs.~\eqref{xi_y_sol} and \eqref{xi_z_sol},
\begin{align}
\label{xi_r_withH}\xi_r&=\frac{r_0^2}{2(r_0-2M)}\frac1\rho\left(\frac{\mu(1-v^2)^{-1/2}}{\sqrt{1-v^2\sin^2\alpha}}\right.\\
\nonumber&\qquad\qquad-\cos(2\alpha)(H+H_{,\alpha\alpha})\bigg)+O(\alpha)+O(\rho),
\end{align}
and the term involving \(H(\alpha)\) reduces to,
\begin{align*}
H+H_{,\alpha\alpha}&=\sum_{n=-\infty}^{\infty}(1-n^2)b_ne^{in\alpha},\\
&=\sum_{n=-\infty}^{\infty}af_ne^{in\alpha},\\
&=-\frac{2\mu\mathcal{L}^2(1-v^2)^{1/2}}{r_0^2\sqrt{1-v^2\sin^2\alpha}},
\end{align*}
using Eq.~\eqref{BOfouriermodes} in the second line and the definition of \(f_n\) in the third line. After substitution into Eq.~\eqref{xi_r_withH}, we are left with,
\begin{align}
\xi_r&=\frac{\mu r_0^2}{2(r_0-2M)}\frac{(1-v^2)^{1/2}}{\rho\sqrt{1-v^2\sin^2\alpha}}\\
\nonumber&\quad\times\left(\frac{r_0-4M\sin^2\alpha}{r_0-3M}\right)+O(\alpha)+O(\rho),
\end{align}
where we have used the value of the specific angular momentum for a circular orbit, Eq.~\eqref{energymomentum}. The gauge vector \(\xi_r\) vanishes for select values of \(\alpha\) when the orbit is within \(r_0\le 4M\), but the \(1/s\) singularity in \(\xi_r\) is entirely unavoidable for any physical circular orbit \(r_0>4M\), and \(\xi_r\) does not satisfy the sufficiently regular criterion for a gauge transformation. This result motivates the Locally Lorenz gauge regularization used in Sec.~\ref{regularization}.

\pagebreak

\end{appendices}

\bibliography{citations}

\end{document}